\documentclass[10pt,oneside,english,traditabstract]{amsart}
\usepackage{mathptmx}

\usepackage[T1]{fontenc}
\usepackage[latin9]{inputenc}
\usepackage{babel}
\usepackage{array}
\usepackage{rotfloat}
\usepackage{url}
\usepackage{amsthm}
\usepackage{amssymb}
\usepackage{graphicx}
\usepackage[authoryear]{natbib}
\usepackage[unicode=true,
 bookmarks=true,bookmarksnumbered=false,bookmarksopen=false,
 breaklinks=false,pdfborder={0 0 0},backref=false,colorlinks=false]
 {hyperref}
\hypersetup{pdftitle={Lutetia},
 pdfauthor={Pedro H. Hasselmann},
 pdfkeywords={lutetia},
 linkcolor=magenta, urlcolor=blue, citecolor=blue}

\makeatletter

\providecommand{\tabularnewline}{\\}

\numberwithin{equation}{section}
\numberwithin{figure}{section}
\newenvironment{lyxlist}[1]
{\begin{list}{}
{\settowidth{\labelwidth}{#1}
 \setlength{\leftmargin}{\labelwidth}
 \addtolength{\leftmargin}{\labelsep}
 }}
{\end{list}}

\@ifundefined{showcaptionsetup}{}{%
 \PassOptionsToPackage{caption=false}{subfig}}
\usepackage{subfig}
\makeatother

\begin{document}

\title{Asteroid (21) Lutetia: Disk-resolved Photometric Analysis of Baetica
Region}

\author{P. H. Hasselmann\textsuperscript{1,2} \and M. A. Barucci\textsuperscript{2}
\and S. Fornasier\textsuperscript{2} \and C. Leyrat\textsuperscript{2}
\and J. M. Carvano\textsuperscript{1} \and D. Lazzaro\textsuperscript{1}
\and H. Sierks\textsuperscript{3}}

\maketitle
\ 

\textsuperscript{1}Observat�rio Nacional (COAA), Rua General Jos�
Cristino 77, S�o Crist�v�o, CEP20921-400, Rio de Janeiro RJ, Brazil. 

\textsuperscript{2}LESIA, Observatoire de Paris, PSL Research University,
CNRS, Univ. Paris Diderot, Sorbonne Paris Cit�, UPMC Univ. Paris 06,
Sorbonne Universit�s, 5 Place J. Janssen, 92195 Meudon Principal Cedex,
France

\textsuperscript{3}Max Planck Institute for Solar System Research,
Justus-von-Liebig-Weg 3, 37077, G�ttingen, Germany

\ 

\begin{center}
corresponding author at: hasselmann@on.br
\par\end{center}

\ 
\begin{abstract}
(21) Lutetia has been visited by Rosetta mission on July 2010 and
observed with a phase angle ranging from 0.15 to 156.8 degrees. The
Baetica region, located at the north pole has been extensively observed
by OSIRIS cameras system. Baetica encompass a region called North
Pole Crater Cluster (NPCC), shows a cluster of superposed craters
which presents signs of variegation at the small phase angle images.
For studying the location, we used 187 images distributed throughout
14 filter recorded by the NAC (Narrow Angle Camera) and WAC (Wide
Angle Camera) of the OSIRIS system on-board Rosetta taken during the
fly-by. Then, we photometrically modeled the region using Minnaert
disk-function and Akimov phase function to obtain a resolved spectral
slope map at phase angles of $5^{\circ}$ and $20^{\circ}$. We observed
a dicothomy between Gallicum and Danuvius-Sarnus Labes in the NPCC,
but no significant phase reddening ($-0.04\pm0.045\%\cdot microns^{-1}deg^{-1}$).
In the next step, we applied the Hapke (2008,2012) model for the NAC
F82+F22 (649.2 nm), WAC F13 (375 nm) and WAC F17 (631.6 nm) and we
obtained normal albedo maps and Hapke parameter maps for NAC F82+F22.
On Baetica, at 649.2 nm, the geometric albedo is $0.205\pm0.005$,
the average single-scattering albedo is $0.181\pm0.005$, the average
asymmetric factor is $-0.342\pm0.003$, the average shadow-hiding
opposition effect amplitude and width are $0.824\pm0.002$ and $0.040\pm0.0007$,
the average roughness slope is $11.45^{\circ}\pm3^{\circ}$ and the
average porosity is $0.85\pm0.002$. We are unable to confirm the
presence of coherent-backscattering mechanism. In the NPCC, the normal
albedo variegation among the craters walls reach 8\% brighter for
Gallicum Labes and 2\% fainter for Danuvius Labes. The Hapke parameter
maps also show a dicothomy at the opposition effect coefficients,
single-scattering albedo and asymmetric factor, that may be attributed
to the maturation degree of the regolith or to compositonal variation.
In addition, we compared the Hapke (2008, 2012) and Hapke (1993) parameters
with laboratory samples and other small-solar system bodies visited
by space missions.
\end{abstract}


\newpage{}

\section{Introduction}

The asteroid (21) Lutetia was observed during a flyby by Rosetta mission
on July 10th, 2010. Rosetta is an ESA cornerstone mission, launched
on March 2nd, 2004, composed of two elements: the Orbiter and the
lander Philae, with the aim to visit the comet 67P/Churyumov-Gerasimenko.
Rosetta spacecraft reached the comet from the heliocentric distance
of about 4 AU to start the characterization of the nucleus prior to
the delivery of the Philae lander (November 12, 2014) and followed
it until its perihelion passage in August 2015 with the end of extended
mission on September 2016. 

The final choice of the targets (21) Lutetia and (2867) Steins \citep{2005A&A...430..313B}
was made only after the launch of the mission and the first orbital
correction man�uvre. The two asteroids have been selected for their
high scientific return and Lutetia in particular because of its large
size which was expected to lead to accurate mass and density determinations. 

Several instruments were active, resulting into acquiring images and
spectrometric observations from the ultraviolet (70 nm, by ALICE UV
spectrometer) through the visible (by OSIRIS imaging system) and infrared
(by the VIRTIS imaging spectrometer) to the millimeter range (0.5-1.3
mm by the MIRO microwave spectrometer), and the radio science investigation
(see \citealp{2015ASTIV...BFMT}, for detailed results). 

The Rosetta spacecraft flew by (21) Lutetia obtaining resolved images
for about 10 hours before the closest approach and revealing an object
with a highly complex history. The OSIRIS camera systems \citep{2007Icar..187...87K}
composed of two cameras (NAC, Narrow Angle Camera and WAC, Wide Angle
Camera) observed the asteroid \citep{2011Sci...334..487S} in 21 broad
and narrow band filters covering more than 50\% of the surface with
spatial scales up to 60 m/pixel. The rotational period of 8.168 hours
and direction of the pole axis were improved \citep{2010A&A...523A..94C}.
The global shape with principal axes dimensions of $121\pm1\ km\times101\pm1\ km\times75\pm13\ km$
was determined \citep{2011Sci...334..487S}, even if a large fraction
of the asteroid\textquoteright s southern hemisphere was not visible
during the fly-by and consequently the shortest semi-major dimension
is not well constrained. From the shape model, an estimation of volume
of $(5.0\pm0.4)\cdot10^{5}km^{3}$ has been derived and combining
it with the mass obtained by the Radio science investigation $1.7\cdot10^{18}kg\pm1
$ by \citet{2011Sci...334..491P} a bulk density of $3.4\pm0.3g\cdot cm^{-3}$
has been computed. 

The north pole is located near a depression that has been produced
by multiple impacts called NPCC (North Polar Crater Clusters). The
north rotational pole was roughly pointed towards the Sun at the time
of the Rosetta encounter and hence high-resolution imaging was restricted
by the illumination to one hemisphere. (21) Lutetia shows to have
geological complex surface dominate by impact craters, landslides
and a diversity set of lineaments (see \citealp{2012P&SS...66...96T},
for detailed description). The largest visible depression is the Massilia
structure, which is a highly degraded crater-like structure of 57
km diameter. The surface shows a remarkable structure with boulders
and landslides. Its rim appears to have been modified by subsequent
impacts. Another dominant feature is the NPCC itself, which forms
the most striking structure within the Baetica region. This appears
to be one of the youngest surfaces on the object and results of several
impacts of varying size, which have overlapped each other. 

The relatively low density of smaller impact craters within the NPCC
contrasts sharply with the high crater density seen in other regions.
\citet{2012P&SS...66...87M} studied the crater distribution and estimate
the age for different regions on Lutetia. Significant amounts of ejecta
were also observed on the far side of the impact crater. Detailed
results on Lutetia surface characterization are reported and discussed
by \citet{2015ASTIV...BFMT}.

In this paper we analyze the Baetica region (Figure \ref{fig:baetica})
as observed by the OSIRIS camera. We apply empirical and theorical
photometric analysis to derive the spectral slope map and semi-physical
parameters of the surface. All analysis is done in disk-resolved scale.
The region is our main focus of interest due to the identification
of some talus of bright material coming down though one of the walls
of the NPCC at the small phase angle images \citep{2012P&SS...66....2S,2012P&SS...66...43M,2012P&SS...66...96T},
moreover it is the only one to be most extensively covered during
all fly-by. Therefore, we are interest to investigate possible space
weathering processes \citep{2010Icar..209..564G}, and for this purpose,
we analyze real variegations and parametric variations among Gades,
Corduba and surrouding areas.

\begin{center}
\begin{figure}
\begin{centering}
\includegraphics[scale=0.5]{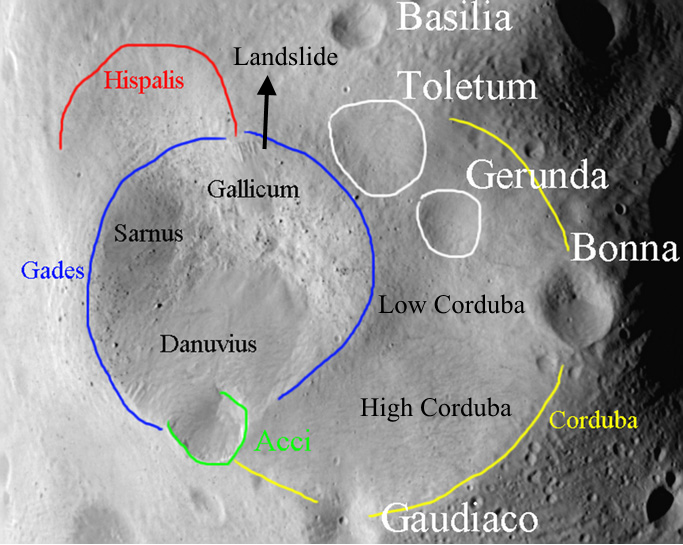}\protect\caption{\label{fig:baetica} The geomorphological units of the Baetica Region
reproduced from \citet{2012P&SS...66...96T}. \textsl{Sarnus Labes,
Gallicum Labes, Danuvius Labes} are the major landslides on NPCC.
Gades has diameter 21 km, Corduba - 34 km, Hispalis - 14 km e Acci
- 6 km. On Corduba, we have separated the labelling in the
Lowland and Highland according to the topographical slope (we reference
Figure 4 in \citealt{2012P&SS...66...87M} for a topographical map).}

\par\end{centering}

\end{figure}

\par\end{center}

(21) Lutetia has been observed with a phase angle ranging from 0.15
to 156.85 degrees, a varying resolution of 0.375 km/pixel to 0.059
km/pixel and limited incidence angle conditions due the fast fly-by
of about 15 km/s. Only a photometric modeling covering all aspects
of the phase curve of an atmosphereless body may provide complete
analysis of the (21) Lutetia surface. The main aspect is the opposition
effect, a non-linear increase in brightness when the phase angles
approaches zero degrees, observed in asteroids for the first time
by \citet{1956AJ.....61..178G}. This has been studied for the connection
with two important optical mechanisms of particulate media: The shadow-hiding
(e.g., \citealp{1981JGR....86.4571H,1990TTSP...19..317L,1999JQSRT..63..445S})
and the coherent-backscattering enhancement (e.g., \citealp{1994IAUS..160..271M,2009ApJ...705L.118M}).
The former is related to the disappearance of mutual shadows among
the regolith particles when opposition is near. The latter happens
when the multiple scattered electromagnetic waves constructively interfere
at near zero phase angle. Those mechanisms are intrinsically connected
to regolith packing, size distribution, particle shape, inclusions
and transparency in a complex relation that it is still subject of
on-going research (e.g., \citealp{2012ApJ...760..118M,2013Icar..226.1465D}).
Other aspect is the macroscopic shadowing, a mechanism often significant
when a surface is observed at phase angle larger than about 30 degrees
and when illumination azimuth is higher. Generally, boulders, micro-craters
or micro-irregularities are evoked to explain the hindering of brightness
due to the casting of large shadows \citep{1984Icar...59...41H,2010Icar..208..548G,2012JQSRT.113.2431S}.

Previous studies on disk-resolved small Solar System bodies have basically
relied on a class of models based on Radiative Transfer Equation (RTE)
to retrieve semi-physical parameters related to those mechanisms.
The Hapke \citeyearpar{1981JGR....86.4571H,1993tres.book.....H,2002Icar..157..523H,2008Icar..195..918H,9781139025683}
models have been the most widespread among the treatment of spacecraft
data of atmosphereless bodies of the Solar System. (433) Eros, orbited
by NEAR Shoemaker \citep{2002Icar..155..189C}, Phobos and Deimos,
visited by Viking Orbiter \citep{1996Icar..123..536T,1998Icar..131...52S}
and (25143) Itokawa, orbited by Hayabusa \citep{2008Icar..194..137K}
and (4) Vesta, orbited by Dawn \citep{2013Icar..226.1252L}, are examples
of small bodies which had their global Hapke parameters obtained.
However, only recently, efforts have been taken to derive a spatially
resolved Hapke parameters out of disk-resolved data \citep{2012Icar..221.1101S,2014JGRE..119.1775S}.

Moving further from the RTE models, there is a class of empirical
models that decouples the reflectance dependence of phase angles,
the phase function, from the gradient behavior due to local topography,
called disk function (e.g., \citealt{2011P&SS...59.1326S}). Generally,
those models neglect or roughly describe the multiple-scattering and
the macroscopic roughness, making their application restricted to
dark smooth surfaces on limited phase angle coverage. A recent application
of such models was undertaken by \citet{2013P&SS...85..198S}, where
classical light scattering laws as Lambert, Minnaert, Lommel-Seeliger
and Akimov were used to photometrically correct images of (4) Vesta,
and an empirical phase function was used to fit each single pixel
on the illuminated surface.

The OSIRIS camera on-board Rosetta observed (21) Lutetia using 23
filters, however the solar phase angle coverage and number of images
is not uniform for all filters. We therefore model the filters following
two different approaches. First, for normalizing images of undersampled
filters to a standard illumination condition, we tested five scattering
laws to find the most suitable one for the photometric correction
of topography, in same manner as \citet{2013P&SS...85..198S}. Since
the phase dependence can vary throughout (21) Lutetia's surface, the
individual fitting of Akimov phase function \citep{1988KFNT....4...10A,2011P&SS...59.1326S}
was obtained for each surface element of the shape model, called facet.
The spectro-photometry of each facet is derived through the normalization
to same a phase angle, allowing a spectral slope map to be obtained.
On the second approach, for the filters NAC F82+F22 (649 nm), WAC
F13 (375 nm) and WAC F17 (630 nm), where the phase angle coverage
was more extensive, we undertake a photometric analysis with Hapke
model \citep{2011Icar..215...83H,9781139025683} to determine photometric
parameters at each facet and thus got their albedo, porosity, macroscopic
roughness and opposition surge. Hence, we expect modeling the opposition
effect at better precision and verify if the albedo variations at
the Baetica regions are connected with different soil properties.

Lastly, on the following sections of this work, we describe the characteristics
of data and shape model used on this article (Section 2.1), we specify
how the emergence, incidence and azimuth angles were obtained through
an image simulator (Section 2.2), we explain the photometric correction
method (Section 3.1), we present the color and spectral slope maps
obtained through the method (Section 3.2) and the locally-resolved
Hapke parameters (Section 4). We then conclude and interpret results
and their implication on the nature of Baetica albedo variations (Section
5 and 6).

\section{Observations and methods}

\subsection{\textmd{OSIRIS Images and Lutetia shape model}}

Resolved images of (21) Lutetia started to be taken at 9:30 hours
before the closest-approach (CA) at a distance of approximately 500.000
km, reaching some 20 pixels wide. The closest-approach happened at
3168.2 km with a velocity of 15 km/s. Several hours later, clear images
of Baetica region were taken, at 46 minutes before the CA, on a pixel
scale of 0.375 km/pixel and phase angle of 1.9 degrees. From this
point, Baetica is on field-of-view until became a dim limb at 4 minutes
after the CA at phase angle of 126 degrees. Table \ref{tab:OSIRIS-images}
presents the characteristics of the filters and images used in this
work.

The OSIRIS obtained 462 images on 27 filter combinations. 234 images
were taken with NAC, and 228 with WAC. All images used in this work
have quality of level 3 on OSIRIS pipeline calibration, which means
they were corrected to dark current, bias, coherent noise, geometry
distortion and flat field. The images were calibrated to absolute
radiance ($W\cdot m^{-2}\cdot nm^{-1}\cdot sr^{-1}$) and converted
to radiance factor (I/F) by dividing them to the incoming solar irradiance
given by $F=S_{o}/\pi R^{2}$ at Lutetia heliocentric distance of
$R=2.72$ UA and solar spectral irradiance $S_{o}$($W\cdot m^{-2}\cdot nm^{-1}$,
normalized at 1 AU). The solar spectral irradiance is set to the central
wavelength of each filter combination.

The uncertainties of OSIRIS camera change for each filter, mainly
due to uncertainties related to the zero point flux. NAC F16 in the
UV have errors about 4-6\%, while NAC and WAC filters in the visible
and NIR have just 2\% of uncertainties in respect to the absolute
flux. WAC F13, however, have estimated errors of about 8-10\%. The
internal coherence error for each image, i.e., how much noise is able
to mischaracterize the image, are estimated to 1.5-2\% \citep{2012P&SS...66...43M,2015A&A...573A..62T}. 

The shape model for (21) Lutetia was obtained through the stereo-photogrammetric
analysis \citep{2011epsc.conf..776J,2012AGUFM.P23B1937C}. The shape
model has 3145728 facets and 1579014 vertices, comprising a file of
175 MB \citep{2013PDS...F}. Each facet represent a surface element
and has about $0.01\ km^{2}$. Detailed shape models are essential
to undertake a disk-resolved photometric correction.

\begin{center}
\begin{table*}[p]
\protect\caption{\label{tab:OSIRIS-images} Summary of the selected OSIRIS NAC and
WAC images acquired during the Lutetia fly-by.}

\begin{centering}
\begin{tabular}{c>{\centering}m{1.6cm}>{\centering}m{1.3cm}>{\centering}m{2cm}>{\centering}m{1cm}c>{\centering}m{1.4cm}}
\cline{1-6} 
{\footnotesize{}Filter} & {\footnotesize{}central wavelength (nm)} & {\footnotesize{}FWHM (nm)} & {\footnotesize{}Image range (2010-07-10 UT)} & {\footnotesize{}Phase range ($^{\circ}$)} & {\footnotesize{}Number} & {\footnotesize{}Solar}{\footnotesize \par}

{\footnotesize{}spectral irradiance $S_{o}$}\tabularnewline
\hline 
{\scriptsize{}NAC F82+F22} & {\footnotesize{}649.2} & {\footnotesize{}84.5} & {\scriptsize{}15.26.05.175 }{\scriptsize \par}

{\scriptsize{}- }{\scriptsize \par}

{\scriptsize{}15.50.39.219} & {\scriptsize{}0.15}{\scriptsize \par}

{\scriptsize{}-}{\scriptsize \par}

{\scriptsize{}144.15} & {\footnotesize{}47} & {\footnotesize{}1.5650}\tabularnewline
\cline{4-5} 
{\scriptsize{}NAC F83+F23} & {\footnotesize{}535.7} & {\footnotesize{}62.4} & {\scriptsize{}14.45.05.650}{\scriptsize \par}

{\scriptsize{}-}{\scriptsize \par}

{\scriptsize{}15.43.27.562} & {\scriptsize{}3.08}{\scriptsize \par}

{\scriptsize{}-}{\scriptsize \par}

{\scriptsize{}61.0} & {\footnotesize{}5} & {\footnotesize{}1.9950}\tabularnewline
\cline{4-5} 
{\scriptsize{}NAC F84+F24} & {\footnotesize{}480.7} & {\footnotesize{}74.9} & {\scriptsize{}14.44.58.330}{\scriptsize \par}

{\scriptsize{}- }{\scriptsize \par}

{\scriptsize{}15.43.19.535} & {\scriptsize{}2.95}{\scriptsize \par}

{\scriptsize{}-}{\scriptsize \par}

{\scriptsize{}59.03} & {\footnotesize{}11} & {\footnotesize{}2.0600}\tabularnewline
\cline{4-5} 
{\scriptsize{}NAC F88+F28} & {\footnotesize{}743.7} & {\footnotesize{}64.1} & {\scriptsize{}15.15.28.228}{\scriptsize \par}

{\scriptsize{}- }{\scriptsize \par}

{\scriptsize{}15.43.43.881} & {\scriptsize{}3.51}{\scriptsize \par}

{\scriptsize{}-}{\scriptsize \par}

{\scriptsize{}65.1} & {\footnotesize{}7} & {\footnotesize{}1.2890}\tabularnewline
\cline{4-5} 
{\scriptsize{}NAC F16} & {\footnotesize{}360.0} & {\footnotesize{}51.1} & {\scriptsize{}14.44.51.065}{\scriptsize \par}

{\scriptsize{}- }{\scriptsize \par}

{\scriptsize{}15.43.11.261} & {\scriptsize{}2.82}{\scriptsize \par}

{\scriptsize{}-}{\scriptsize \par}

{\scriptsize{}57.07} & {\footnotesize{}7} & {\footnotesize{}1.0305}\tabularnewline
\cline{4-5} 
{\scriptsize{}NAC F41} & {\footnotesize{}882.1} & {\footnotesize{}65.9} & {\scriptsize{}14.45.50.902}{\scriptsize \par}

{\scriptsize{}- }{\scriptsize \par}

{\scriptsize{}15.43.43.881} & {\scriptsize{}3.93}{\scriptsize \par}

{\scriptsize{}-}{\scriptsize \par}

{\scriptsize{}67.15} & {\footnotesize{}7} & {\footnotesize{}0.9230}\tabularnewline
\cline{4-5} 
{\scriptsize{}NAC F51} & {\footnotesize{}805.3} & {\footnotesize{}40.5} & {\scriptsize{}14.45.43.923}{\scriptsize \par}

{\scriptsize{}-}{\scriptsize \par}

{\scriptsize{}15.48.45.539} & {\scriptsize{}4.15}{\scriptsize \par}

{\scriptsize{}-}{\scriptsize \par}

{\scriptsize{}128.65} & {\footnotesize{}5} & {\footnotesize{}1.1180}\tabularnewline
\cline{4-5} 
{\scriptsize{}NAC F61} & {\footnotesize{}931.9} & {\footnotesize{}34.9} & {\scriptsize{}14.45.58.036}{\scriptsize \par}

{\scriptsize{}-}{\scriptsize \par}

{\scriptsize{}15.49.01.668} & {\scriptsize{}4.01}{\scriptsize \par}

{\scriptsize{}-}{\scriptsize \par}

{\scriptsize{}130.41} & {\footnotesize{}5} & {\footnotesize{}0.8480}\tabularnewline
\cline{4-5} 
{\scriptsize{}NAC F71} & {\footnotesize{}989.3} & {\footnotesize{}38.2} & {\scriptsize{}14.46.05.075}{\scriptsize \par}

{\scriptsize{}-}{\scriptsize \par}

{\scriptsize{}15.44.00.348} & {\scriptsize{}4.22}{\scriptsize \par}

{\scriptsize{}-}{\scriptsize \par}

{\scriptsize{}69.36} & {\footnotesize{}9} & {\footnotesize{}0.7363}\tabularnewline
\cline{4-5} 
{\scriptsize{}WAC F13} & {\footnotesize{}375.6} & {\footnotesize{}9.8} & {\scriptsize{}15.15.20.596}{\scriptsize \par}

{\scriptsize{}-}{\scriptsize \par}

{\scriptsize{}15.47.49.781} & {\scriptsize{}0.15}{\scriptsize \par}

{\scriptsize{}-}{\scriptsize \par}

{\scriptsize{}121.44} & {\footnotesize{}30} & {\footnotesize{}1.1030}\tabularnewline
\cline{4-5} 
{\scriptsize{}WAC F15} & {\footnotesize{}572.1} & {\footnotesize{}11.5} & {\scriptsize{}15.15.23.117}{\scriptsize \par}

{\scriptsize{}-}{\scriptsize \par}

{\scriptsize{}15.43.39.468} & {\scriptsize{}3.18}{\scriptsize \par}

{\scriptsize{}-}{\scriptsize \par}

{\scriptsize{}63.95} & {\footnotesize{}5} & {\footnotesize{}1.8280}\tabularnewline
\cline{4-5} 
{\scriptsize{}WAC F16} & {\footnotesize{}590.7} & {\footnotesize{}4.7} & {\scriptsize{}15.15.25.358}{\scriptsize \par}

{\scriptsize{}-}{\scriptsize \par}

{\scriptsize{}15.43.47.143} & {\scriptsize{}3.22}{\scriptsize \par}

{\scriptsize{}-}{\scriptsize \par}

{\scriptsize{}65.91} & {\footnotesize{}5} & {\footnotesize{}1.8180}\tabularnewline
\cline{4-5} 
{\scriptsize{}WAC F17 } & {\footnotesize{}630.5} & {\footnotesize{}4.0} & {\scriptsize{}15.14.41.474}{\scriptsize \par}

{\scriptsize{}-}{\scriptsize \par}

{\scriptsize{}15.54.40.168} & {\scriptsize{}0.33}{\scriptsize \par}

{\scriptsize{}-}{\scriptsize \par}

{\scriptsize{}149.61} & {\footnotesize{}39} & {\footnotesize{}1.6300}\tabularnewline
\cline{4-5} 
{\scriptsize{}WAC F18} & {\footnotesize{}612.6} & {\footnotesize{}9.8} & {\scriptsize{}15.15.28.330}{\scriptsize \par}

{\scriptsize{}-}{\scriptsize \par}

{\scriptsize{}15.43.55.127} & {\scriptsize{}3.26}{\scriptsize \par}

{\scriptsize{}-}{\scriptsize \par}

{\scriptsize{}68.0} & {\footnotesize{}5} & {\footnotesize{}1.7090}\tabularnewline
\cline{4-5} 
 &  &  &  &  &  & \tabularnewline
\hline 
\end{tabular}
\par\end{centering}

\end{table*}

\par\end{center}

\subsection{\textmd{Retrieving the Photometric Angles}}

The photometric angles are obtained through the OASIS simulator \citep{2010P&SS...58.1097L,2012Icar..221.1101S}
developed by L. Jorda of the Laboratoire d'Astrophysique de Marseille.
The simulator allows us to recreate absolute flux images of a body
on same observational conditions based on a given reflectance model.
For irregular shapes, ray-tracing code reconstructs shadows over the
surface according to incidence angle. The Figure \ref{fig:shape-model}
exemplifies the degree of reproduction of OASIS simulator when rendering
a image using Lambert law.

\begin{center}
\begin{figure}
\begin{centering}
\subfloat[]{\protect\begin{centering}
\protect\includegraphics[scale=0.1]{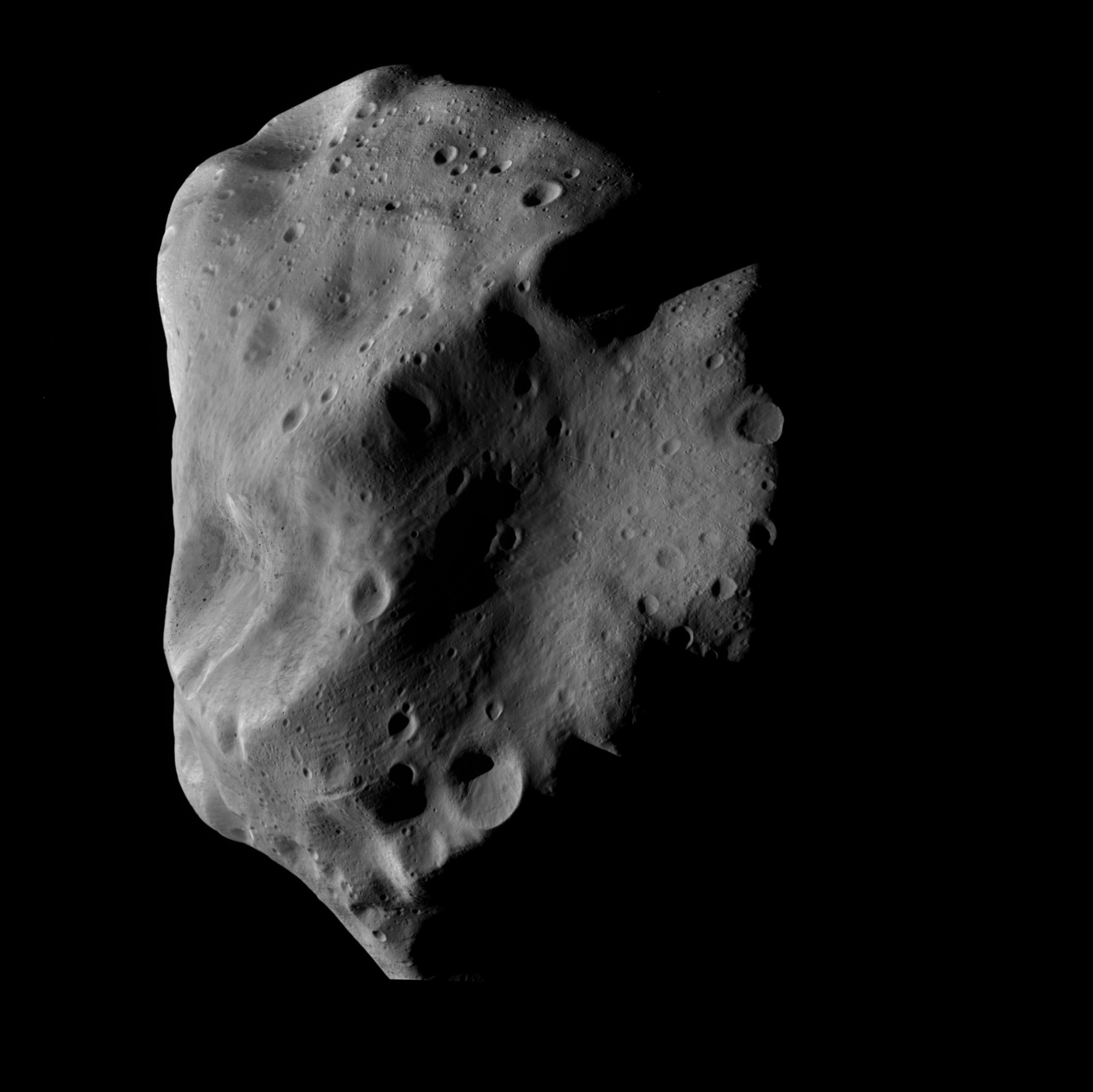}\protect
\par\end{centering}

}\subfloat[]{\protect\centering{}\protect\includegraphics[scale=0.1]{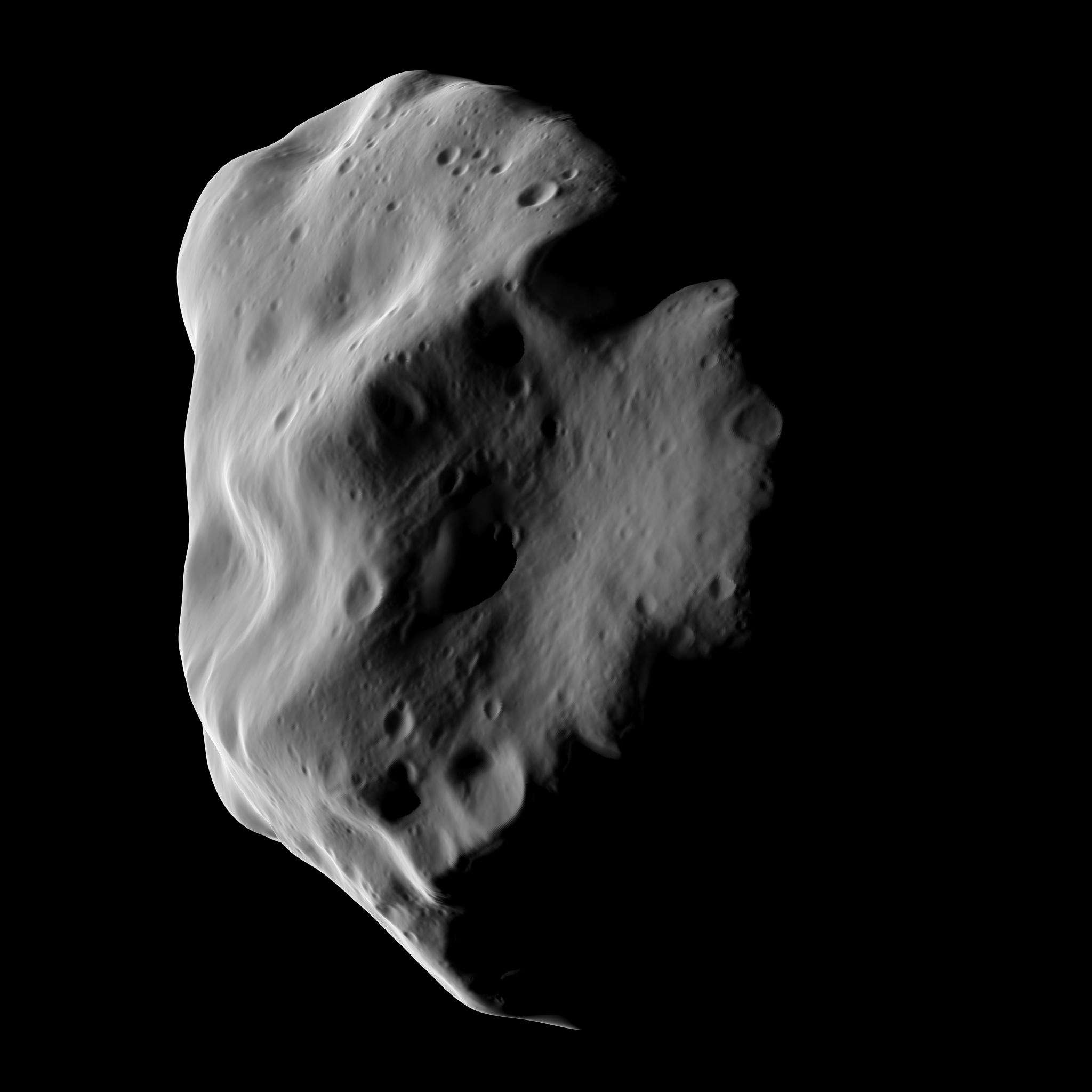}\protect}
\par\end{centering}

\protect\caption{\label{fig:shape-model}(a) Raw image at 15:45:09:210 UT. (b) Rendered
image at same time instant by OASIS using Lambert law to recreate
the surface reflectance.}
\end{figure}

\par\end{center}

The OASIS takes the following information as input:
\begin{itemize}
\item Schedule, \emph{binning}, filter and image size.
\item Rotational period and orientation of the rotational axis.
\item Target and spacecraft trajectory.
\item Spacecraft orientation and instrument placement. 
\item Spacecraft instrument setups as field-of-view, size, gain, coherent
noise and pointing.
\item Target's shape model.
\end{itemize}
Then, it provides:
\begin{itemize}
\item A rendered image of the object on a specific distance and view;
\item a table containing the relation of facets and incidence ($i$), emergence
($e$) and azimuth ($\phi$) angles for the given orbital and rotational
configuration;
\item a table containing intersection of pixels and facets. The fraction
of solid angle $\varOmega_{j}$ carried by the facet in one pixel
or more is also given.
\end{itemize}
The trajetory and instrumental information for Rosetta and (21) Lutetia
are available in the NAIF SPICE kernels\footnote{\url{http://www.cosmos.esa.int/web/spice/spice-for-rosetta}}.
SPICE kernels are continuously updated on the most accurate position
and pointing of the spacecraft as the mission unveils.  However,
the camera off-set has not been incorporated into instrument pointing,
causing shift between real and rendered images, not allowing a perfect
link of facets, measured radiance factor and photometric angles. This
complication is solved through semi-automatic corregistration, to
achieve a sub-pixel overlap in the studied region. An algorithm was
written in Python 2.7.5, using the scikit-image package\footnote{\url{http://scikit-image.org/}},
for such kind of procedure. The user must define at least four control
points on exactly same features observed at the real and rendered
images, these points are then used to calculate a correspondence matrix
used in an affine transformation that shift, rotate and scale the
real image to the rendered image. We observe a degraded matching of
the pixels in the border, therefore all facets in the limb ( $i>80^{\circ}$
and $e>80^{\circ}$) are removed of the analysis.

Thereafter, the OASIS simulator table containing intersection of pixels
and facets allows the calculation of radiance factor for each observed
facet in the asteroid surface. Those values are then computed through
the relation:

\begin{equation}
(I/F)_{k}=\sum_{j}^{n}(I/F)_{j}\frac{\varOmega_{j}}{\varOmega_{k}}
\end{equation}

\noindent where the radiance factor of a given facet $(I/F)_{k}$
is given by the sum of all intersected pixels $j$ weighted by the
ratio of the partial solid angle $\varOmega_{j}$ over the total solid
angle $\varOmega_{k}$ of the facet. Thereby, we end with a full catalog
of photometric angles and radiance factors for each facet.

\begin{center}
\begin{figure*}[p]
\begin{centering}
\subfloat[]{\protect\begin{centering}
\protect\includegraphics[scale=0.35]{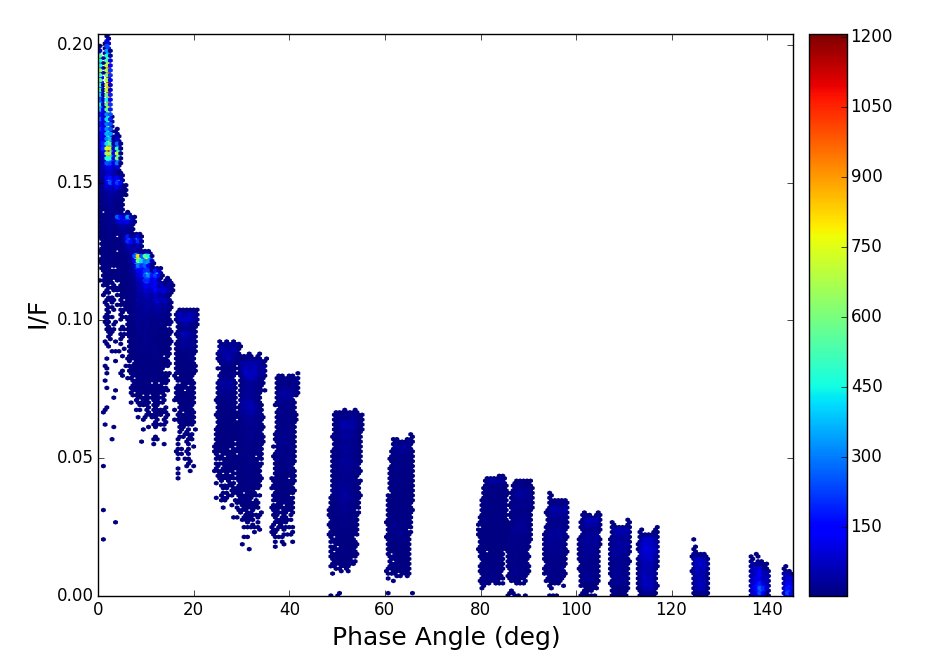}\protect
\par\end{centering}

}\subfloat[]{\protect\begin{centering}
\protect\includegraphics[scale=0.25]{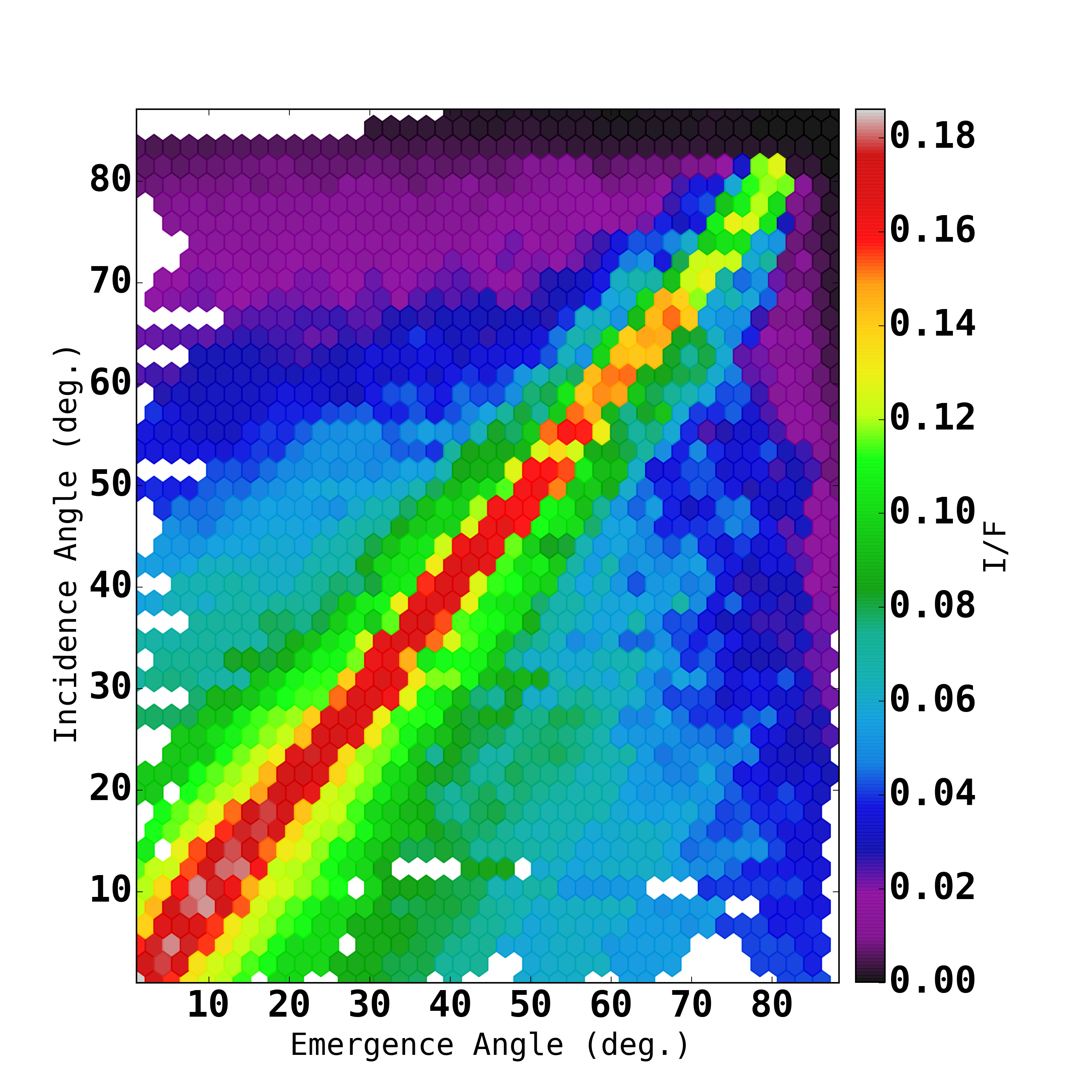}\protect
\par\end{centering}

}
\par\end{centering}

\begin{centering}
\subfloat[]{\protect\begin{centering}
\protect\includegraphics[scale=0.25]{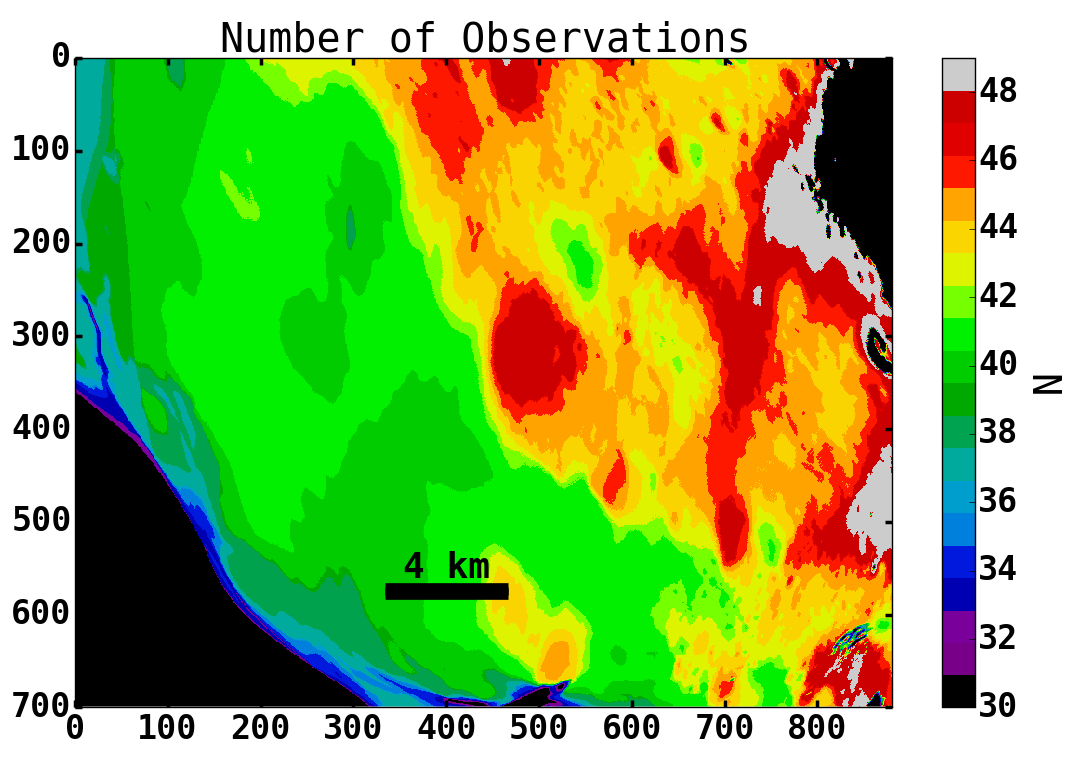}\protect
\par\end{centering}

}\subfloat[]{\protect\begin{centering}
\protect\includegraphics[scale=0.25]{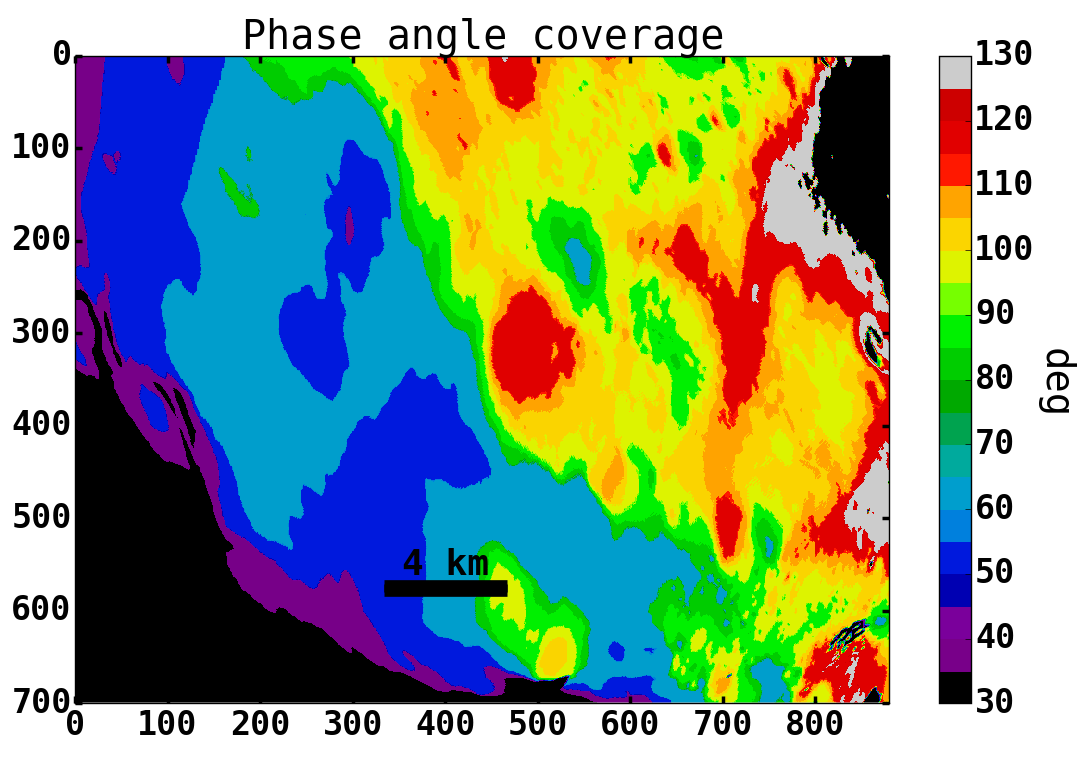}\protect
\par\end{centering}

}
\par\end{centering}

\begin{centering}
\subfloat[]{\protect\begin{centering}
\protect\includegraphics[scale=0.25]{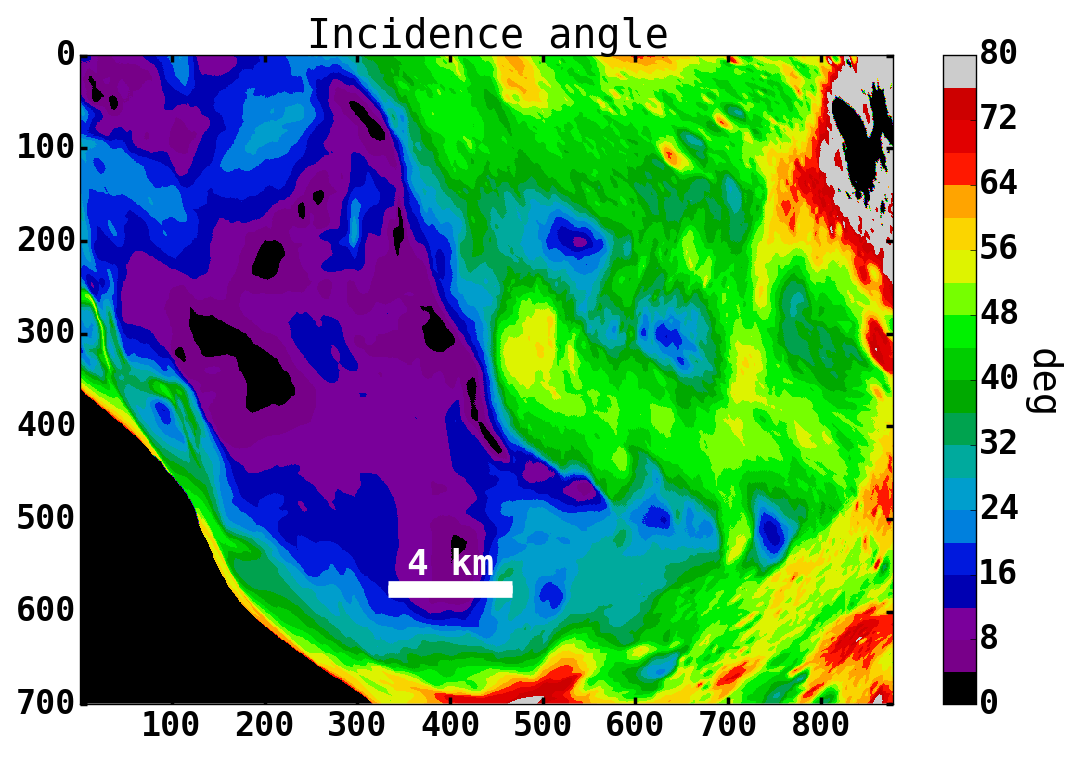}\protect
\par\end{centering}

}
\par\end{centering}

\protect\caption{\label{fig:phot_angles} Radiance factor of (21) Lutetia in function
of the illumination angles. The data was binned on a grid of $40\times40\times40$
(incidence, emergence, azimuth), with 2 degrees of interval on average.
(a) Phase curve of (21) Lutetia in the NAC F82+F22. The color code
indicate the number of measurements. (b) Radiance factor as function
of $i$ and $e$. Map of (c) the number of observations, (d) the phase
angle coverage and (e) incidence angle at Baetica
and Etruria regions for $i<80^{\circ}$ and $e<80^{\circ}$ .}

\end{figure*}

\par\end{center}

The resolved phase curve of Lutetia acquired with NAC F82+F22 is reported
in the Figure \ref{fig:phot_angles}a. Most of the data have been
acquired at relatively small phase angles, thus we have a good sampling
of the opposition surge. If we fit a straight line for the radiance
factor of facets observed with phase angle larger than $20^{\circ}$,
we measure a increase of the radiance factor of 0.11 due to the non-linear
enhancement, which is $\sim50$\% of the albedo at zero phase angle
($0.19\pm0.02$, the geometric albedo according to \citealp{2011Sci...334..487S}).
The spread of points is due to geometric condition only, the so-called
disk profile \citep{2011P&SS...59.1326S}, were facets close to limb
are obscured in the extreme values of incidence and emergence angles.
On Figure \ref{fig:phot_angles}b, we observe a concentration of radiance
factor along diagonal that results from the specular behavior of the
opposition effect, and outside of it, the diffusive scattering component.
Figures \ref{fig:phot_angles}cd shows that most of Etruria and Baetica
areas satisfy the minimum phase angle coverage required to limit the
degeneracy among all Hapke parameters ($>50^{\circ}$, \citealp{Fernando2015271}).
Hapke model analysis is reported in the section 4.

\section{Spectral Slope}

\subsection{\textmd{Photometric Correction}}

For the photometric correction, we relied on models where the explicit
dependence of phase angles is decoupled from topographic light scattering
behavior. Generally, the reliance of this type of modeling is constrained
up to phase angle of $\sim70^{\circ}$. We are aware that resolution
per pixel widely varies across the image, and therefore also varies
the quality and the amount of variegation we might detect in respect
to phase angle.

Radiance factor or apparent albedo is the bi-directional reflectance
$r$ depending at arbitrary illumination conditions:

\begin{equation}
I/F=A_{p}(i,e,\alpha,\lambda)=\pi\cdot r_{(i,e,\alpha,\lambda)}
\end{equation}

Then, radiance factor can be separated in two components, the equigonal
albedo and the disk brightness contribution:

\begin{equation}
I/F=A_{eq}(\alpha,\lambda)\cdot D(i,e,\varphi,\overline{\alpha},\lambda)
\end{equation}
\label{disk-phase-eq}

The equigonal albedo $A_{eq}$ describes the solar phase angle dependence
only, or so-called phase curve. $D(i,e,\varphi,\alpha,\lambda)$ describes
brightness distribution over the disk of the resolved object relative
to the mirror point fixed at $\alpha$, and it is known as disk function.
The disk functions we have implemented are Lambert law, Lommel-Seeliger
law \citep{2005JRASC..99...92F}, Minnaert law \citep{1941ApJ....93..403M},
McEwen or Lunar-Lambertian function \citep{1991Icar...92..298M} and
Oren-Nayar function (Oren \& Nayar \citeyear{CAVE_0095}) (Table \ref{tab:disk-equations}).
This component is related to limb darkening, topography and also the
global surface scattering properties. Further mathematical and qualitative
description of photometric definitions was reviewed by \citet{2011P&SS...59.1326S}.
In what follows, the equigonal albedo is also decomposed in two terms:

\begin{equation}
A_{eq}(\alpha,\lambda)=A_{0}(\lambda)f(\alpha)
\end{equation}

\noindent where the Normal Albedo $A_{0}$ is the equigonal albedo
at $\alpha=0$ and $f(\alpha)$ is the phase function normalized to
unity at $\alpha=0$. Retrieving normal albedo is the main goal for
characterizing real albedo or color variations over the object landscape.
The phase function as proposed by Akimov \citep{1988KFNT....4...10A,2011P&SS...59.1326S}
was derived to be applied to a random rough surface area covered by
semi-translucent particles:

\begin{equation}
f(\alpha)=\frac{e^{-\mu_{1}\alpha}+me^{-\mu_{2}\alpha}}{1+m}
\end{equation}

\noindent where $m$ is the amplitude of the opposition surge, $\mu_{1}$
and $\mu_{2}$ are related to width of the opposition surge and surface
roughness, respectively.

The methodology involved two steps: the disk correction and the facet-by-facet
phase curve fitting. First, we took images of phase angle $<90^{\circ}$
and the data was binned into a $40\times40\times40$ cell grid (about
$2^{\circ}\times2^{\circ}\times(0.1^{\circ}-2^{\circ})$ step) of
the $i$, $e$ and $\phi$ angles. The radiance factor was averaged
for each cell. This was done to soft the aspect of any albedo variation
and bad pixel/facet effect on the disk correction. Then, through the
Levenberg-Marquardt minimization\footnote{\url{http://docs.scipy.org/doc/scipy/reference/generated/scipy.optimize.leastsq.html}}
\citep{1944QAM...2.164L,MINPACK} of the unweighed RMS between measured
I/F and equation 3.2, we obtain the $\bar{A}_{eq}(\alpha,\lambda)$
and a disk coefficient for each image. On the next step, we divide
the disk function to all measured I/F for each image and each facet
to obtain $A_{eq}(\alpha,\lambda)$. On the equigonal albedo, we fit
each facet using again the Levenberg-Marquardt minimization to obtain
the $A_{0}(\lambda)$ and the four parameters of Akimov phase function.
We obtain a table contain Normal Albedo and Akimov parameters for
each facet and the disk coefficient for each image. The operation
is applied for all filters in Table \ref{tab:OSIRIS-images}. The
table of Akimov parameters and Normal Albedos can be replaced in equation
3.3 and 3.4 to reconstruct the Equigonal Albedo for any phase angle.

\begin{center}
\begin{table*}
\protect\caption{\label{tab:disk-equations}Disk functions}

\begin{centering}
\begin{tabular}{ccc}
\hline 
{\small{}Function} & {\small{}Description} & {\small{}Coefficient meaning}\tabularnewline
\hline 
\hline 
{\small{}$D_{LS}=2\frac{\cos(i)}{\cos(i)+\cos(e)}$} & {\small{}Lommel-Seeliger law.} & {\small{}-}\tabularnewline
{\small{}$D_{L}=\cos(i)$} & {\small{}Lambert law.} & {\small{}-}\tabularnewline
{\small{}$D_{LL}=2c\frac{\cos(i)}{\cos(i)+\cos(e)}+(1-c)\cos(i)$} & {\small{}Lunar-Lambertian (McEwen) law.} & {\small{}L-S to Lambert ratio.}\tabularnewline
{\small{}$D_{M}=\cos(i)^{k}\cos(e)^{k-1}$} & {\small{}Minnaert law.} & {\small{}Diversion from lambert law.}\tabularnewline
{\small{}$D_{ON}^{\dagger}$} & Oren-Nayar model. & {\small{}Average spike slope.}\tabularnewline
\hline 
\end{tabular}
\par\end{centering}

\begin{centering}
{\footnotesize{}$^{\dagger}$The expression is cumbersome to be written
here. We invite readers to check Oren \& Nayar\citeyearpar{CAVE_0095}.}
\par\end{centering}{\footnotesize \par}

\end{table*}

\par\end{center}

\subsection{\textmd{Projected map}}

\begin{center}
\begin{figure*}[tp]
\begin{centering}
\includegraphics[scale=0.3]{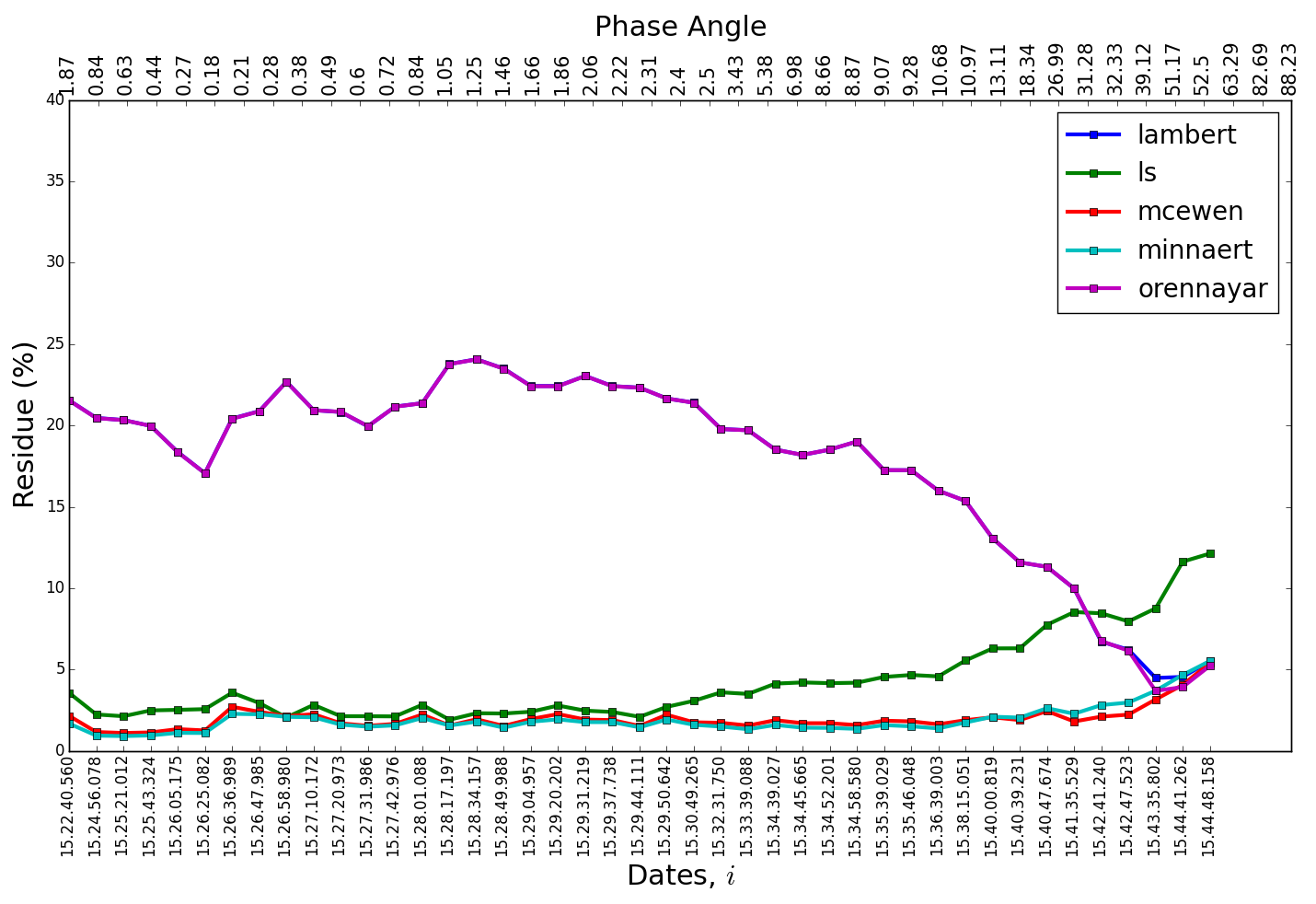}
\par\end{centering}

\protect\caption{\label{fig:disk-residue} The normalized RMS (\%) between the modeled
and measured radiance factor of each image $i$. The residues of Lunar-Lambertian
and Minnaert laws are under 5\%. }

\end{figure*}

\par\end{center}

The disk functions (Table \ref{tab:disk-equations}) were tested on
the NAC F82+F22 images to select the most suitable disk function for
describing the photometric profile of Lutetia. As reported in \ref{fig:disk-residue},
the Lunar-Lambertian and Minnaert laws have the smallest residues.
Both laws have a hybrid behavior, predicting an uniform disk brightness
at zero phase angle and have the Lambert law as limit case ($c=k=1$).
On Figure \ref{fig:disk-coefs}, we present the dependence of the
disk coefficients of each law as function of phase angle. All coefficients
show the same trend: the disk behavior becomes more lambertian as
the phase angle increases, as similar in the Moon \citep{1996LPI....27..841M},
Vesta \citep{2013P&SS...85..198S} and Tempel 1 \citep{2013Icar..222..559L}.
Lutetia presents a steep Minnaert coefficient slope of $0.005/deg$,
similar to Vesta, while dark objects like Tempel 1 present a slope
of $0.002/deg$. Moreover, Lutetia also converges to same Minnaert
coefficient as the Moon, Vesta and Temple 1, at $0.5505\pm0.01$.
We find similar behavior for Lunar-Lambertian law. Such behavior is
interpreted as weak limb darkening at opposition evolving to a bright
limb trend, solely dominated by single scattering. Therefore, we decided
to relay on the Minnaert law for the photometric correction of the
images of the remaining filters.

\begin{center}
\begin{figure}[th]
\begin{centering}
\includegraphics[scale=0.5]{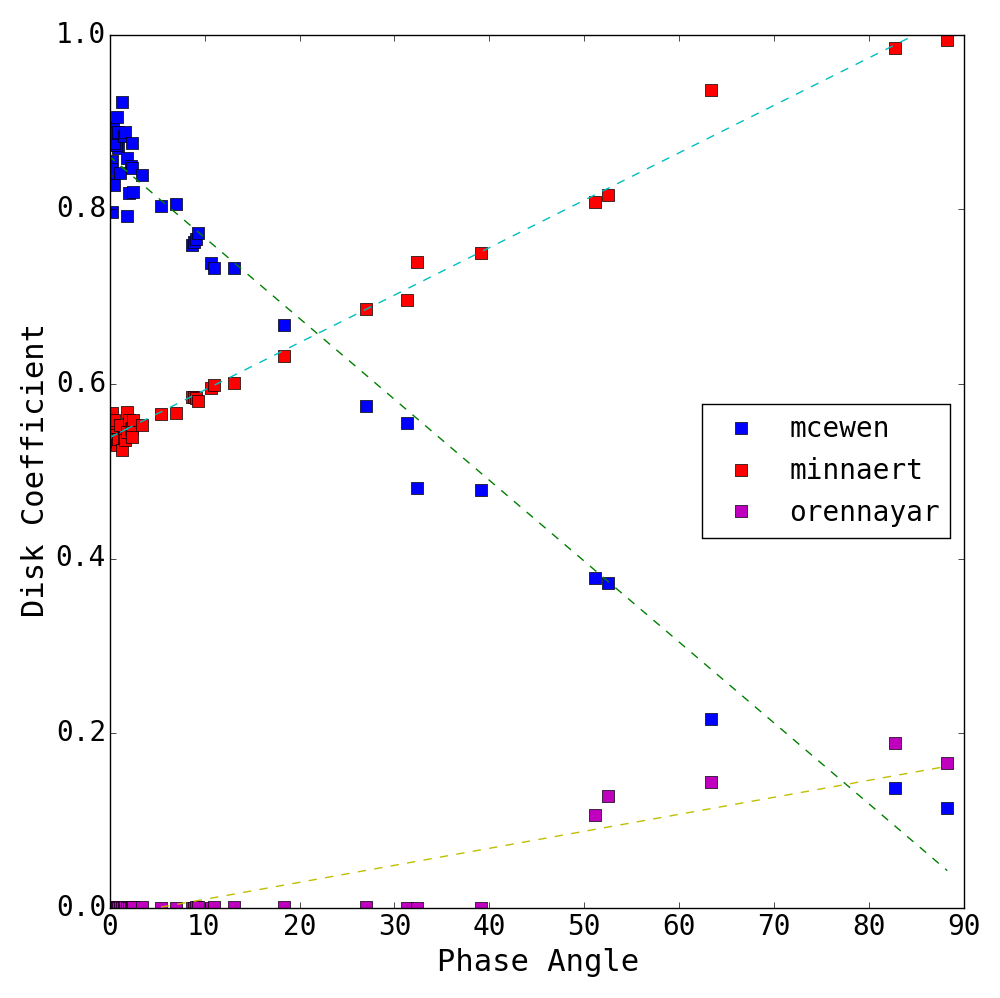}\protect\caption{\label{fig:disk-coefs} Disk coefficients of McEwen (Lunar-Lambettian),
Minnaert and Oren-Nayar laws as function of phase angles.}

\par\end{centering}

\end{figure}

\par\end{center}

The $A_{eq}(\alpha,\lambda)$ for each image and the phase curves
for each filter are illustrated at Figure \ref{fig:phase-curves}.
The Akimov phase parameters are presented in Table \ref{tab:Akimov-phase-parameters}.
Aside NAC F82+22, WAC F13 and WAC F17, the other filters have a poorer
phase angle coverage, and in particular there are no data for phase
angle $<3^{\circ}$ and $>60^{\circ}$, and have quite sparse points.
Therefore we have fixed the parameter $\mu_{1}$ and hence $A_{0}(\lambda)$
becomes better constrained. All filters have most of their images
around phase angles of $5^{\circ}$ and $20^{\circ}$, thus we are
able to obtain a better phase curve fitting on these points. We calculated
the spectrophotometry $A_{eq}(5^{\circ},\lambda)$ and $A_{eq}(20^{\circ},\lambda)$
for each filter and a linear equation was fit to each pixel and its
spectral slope $\gamma(\alpha)$ computed. We then constructed a disk-resolved
projected map of spectral slope $\gamma(\alpha)$ for Baetica region
(Figure \ref{fig:spectral-slope}). Based on the internal coherence
error and an uncertainty on the $i$, $e$ and $\phi$ angles of 0.5
degrees, we estimate a propagated error to $A_{eq}$ of , at most,
6\%. This value is consistent with the variations among neighbouring
pixels and previous estimations with OASIS simulator (5\%, \citealp{2012Icar..221.1101S}).

\begin{center}
\begin{table}
\protect\caption{\label{tab:Akimov-phase-parameters} Global Akimov phase parameters. }

\centering{}%
\begin{tabular}{c>{\centering}p{1cm}c>{\centering}p{1cm}c>{\centering}p{1cm}}
{\footnotesize{}Filter} & {\footnotesize{}$A_{0}$} & {\footnotesize{}$m$} & {\footnotesize{}$\mu_{1}$} & {\footnotesize{}$\mu_{2}$} & \tabularnewline
\hline 
{\small{}NAC F82+F22} & {\small{}0.195} & {\small{}1.32} & {\small{}8.55} & {\small{}0.678} & \tabularnewline
{\small{}NAC F83+F23} & {\small{}0.179} & {\small{}2.02} & {\small{}-} & {\small{}0.861} & \tabularnewline
{\small{}NAC F84+F24} & {\small{}0.176} & {\small{}2.43} & {\small{}-} & {\small{}0.925} & \tabularnewline
{\small{}NAC F88+F28} & {\small{}0.181} & {\small{}2.76} & {\small{}-} & {\small{}0.857} & \tabularnewline
{\small{}NAC F16} & {\small{}0.165} & {\small{}2.28} & {\small{}-} & {\small{}0.902} & \tabularnewline
{\small{}NAC F41} & {\small{}0.187} & {\small{}2.39} & {\small{}-} & {\small{}0.783} & \tabularnewline
{\small{}NAC F51} & {\small{}0.180} & {\small{}2.86} & {\small{}-} & {\small{}0.849} & \tabularnewline
{\small{}NAC F61} & {\small{}0.200} & {\small{}1.39} & {\small{}-} & {\small{}0.365} & \tabularnewline
{\small{}NAC F71} & {\small{}0.203} & {\small{}2.09} & {\small{}-} & {\small{}0.716} & \tabularnewline
{\small{}WAC F13} & {\small{}0.187} & {\small{}1.62} & {\small{}-} & {\small{}0.791} & \tabularnewline
{\small{}WAC F15} & {\small{}0.175} & {\small{}2.70} & {\small{}-} & {\small{}0.871} & \tabularnewline
{\small{}WAC F16} & {\small{}0.177} & {\small{}2.40} & {\small{}-} & {\small{}0.880} & \tabularnewline
{\small{}WAC F17 } & {\small{}0.190} & {\small{}1.69} & {\small{}-} & {\small{}0.768} & \tabularnewline
{\small{}WAC F18} & {\small{}0.174} & {\small{}2.66} & {\small{}-} & {\small{}0.828} & \tabularnewline
 &  &  &  &  & \tabularnewline
\end{tabular}
\end{table}

\par\end{center}

\begin{center}
\begin{figure}[h]
\begin{centering}
\includegraphics[scale=0.4]{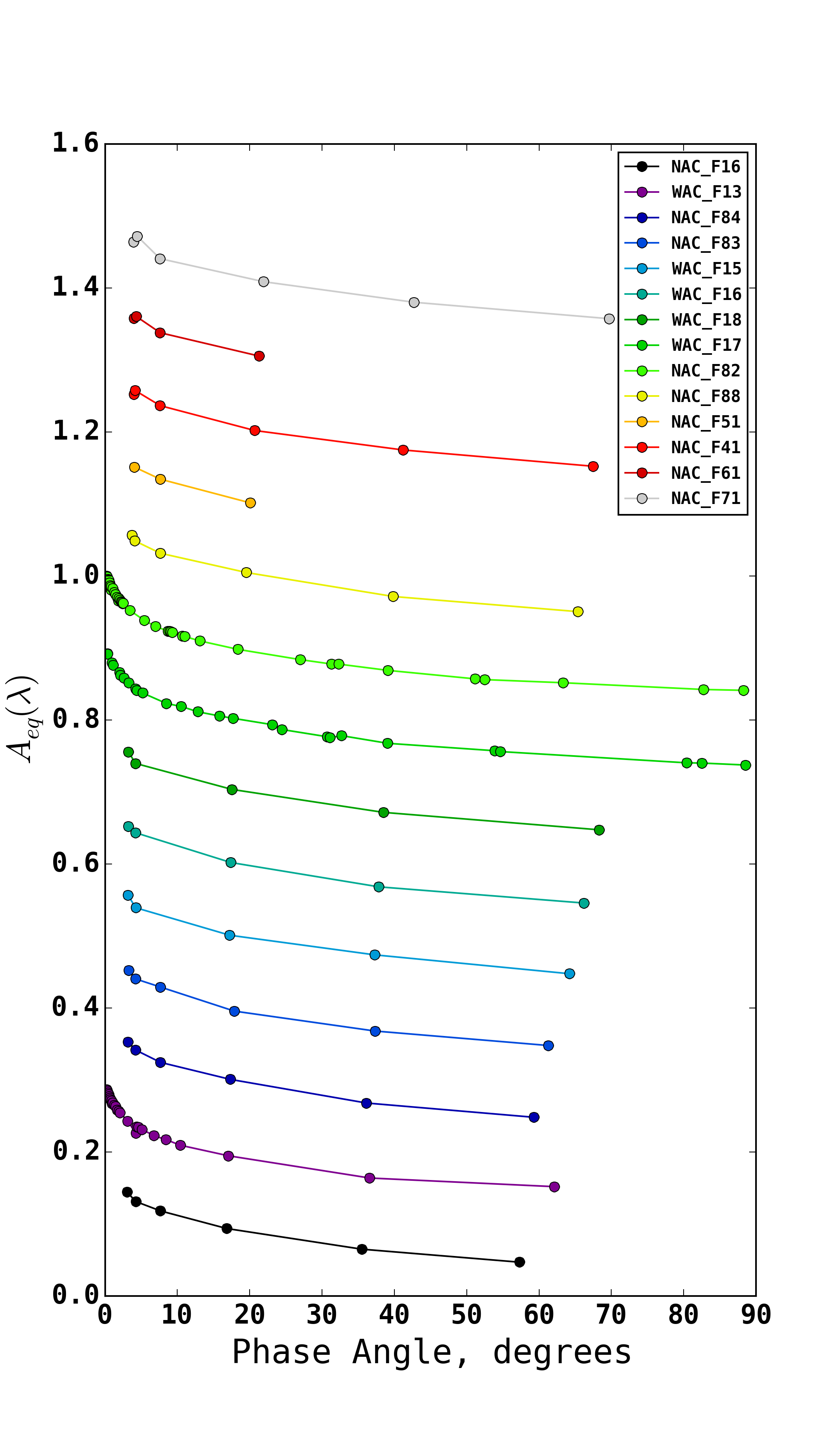}
\par\end{centering}

\protect\caption{\label{fig:phase-curves}The $A_{eq}(\alpha,\lambda)$ for each image
composing a phase curve for each filter. The curves are shifted by
0.1 unit of radiance factor. Only the images at phase angle under
$90^{\circ}$ are used. Over this limit, the photometric correction
must strongly relay on a function to correct the macroscopic shadowing.}

\end{figure}

\par\end{center}

We identify a spectral dichotomy between Gallicum Labes-Low Corduba
and Danuvius Labes-Sarnus Labes-High Corbuda. Gallicum and surroundings
present a spectral slope of about 18-45\% ($4.1-4.8\%\cdot\mu m^{-1}$
) redder than average of $\bar{\gamma}(5^{\circ})=3.5\pm0.55\%\cdot\mu m^{-1}$
at phase angle of $5^{\circ}$, while Danuvius Labes present a neutral
spectrum of slope of about 15-40\% ($2.1-3.0\%\cdot\mu m^{-1}$) bluer
than average. The dichotomy ratio remains similar for the $\gamma(20^{\circ})$-map
($\bar{\gamma}(20^{\circ})=2.9\pm0.38\%\cdot\mu m^{-1}$), but High
Corduba becomes bluer by 17-38\% ($1.8-2.4\%\cdot\mu m^{-1}$). Gallicum
Labes is covered by a redder and brighter material that extends out
of NPCC into the borders of Baetica and Achaia regions. Danuvius Labes,
Sarnus Labes and Corduba, on the other hand, are darker and bluer
and are confined inside NPCC.

The global phase reddening of spectral slope on Baetica is small and
not statistically significant in the considered phase angle range
($\frac{\bar{\gamma}(20^{\circ})-\bar{\gamma}(5^{\circ})}{20^{\circ}-5^{\circ}}=-0.04\pm0.045\%\cdot\mu m^{-1}deg^{-1}$).
However, specifically on Corduba, we observe a small degree of bluing
of $-0.053/-0.066\pm0.045\%\cdot\mu m^{-1}deg^{-1}$, not followed
by the same behavior on Gallicum, Danuvius or Sarnus. Since High Corbuda
and Danuvius Labes are observed at similar incidence and emergence
angles, we might rule out any artifact in the photometric correction. 

Finally, we clearly see a small bluer spot amid Gallicum Labes connected
to the small and smooth landslide we observe in the area (Structure
L, details at \citealp{2012P&SS...66...96T}). The landslide present
the same morphological caracteristics of Danuvius Labes and High Corduba.
Gallicum and Danuvius-Sarnus Labes are landslides of very different
morphological and photometric characteristics that may have been produced
at different temporal events.

\begin{center}
\begin{figure*}[t]
\begin{centering}
\subfloat[]{\protect\begin{centering}
\protect\includegraphics[scale=0.25]{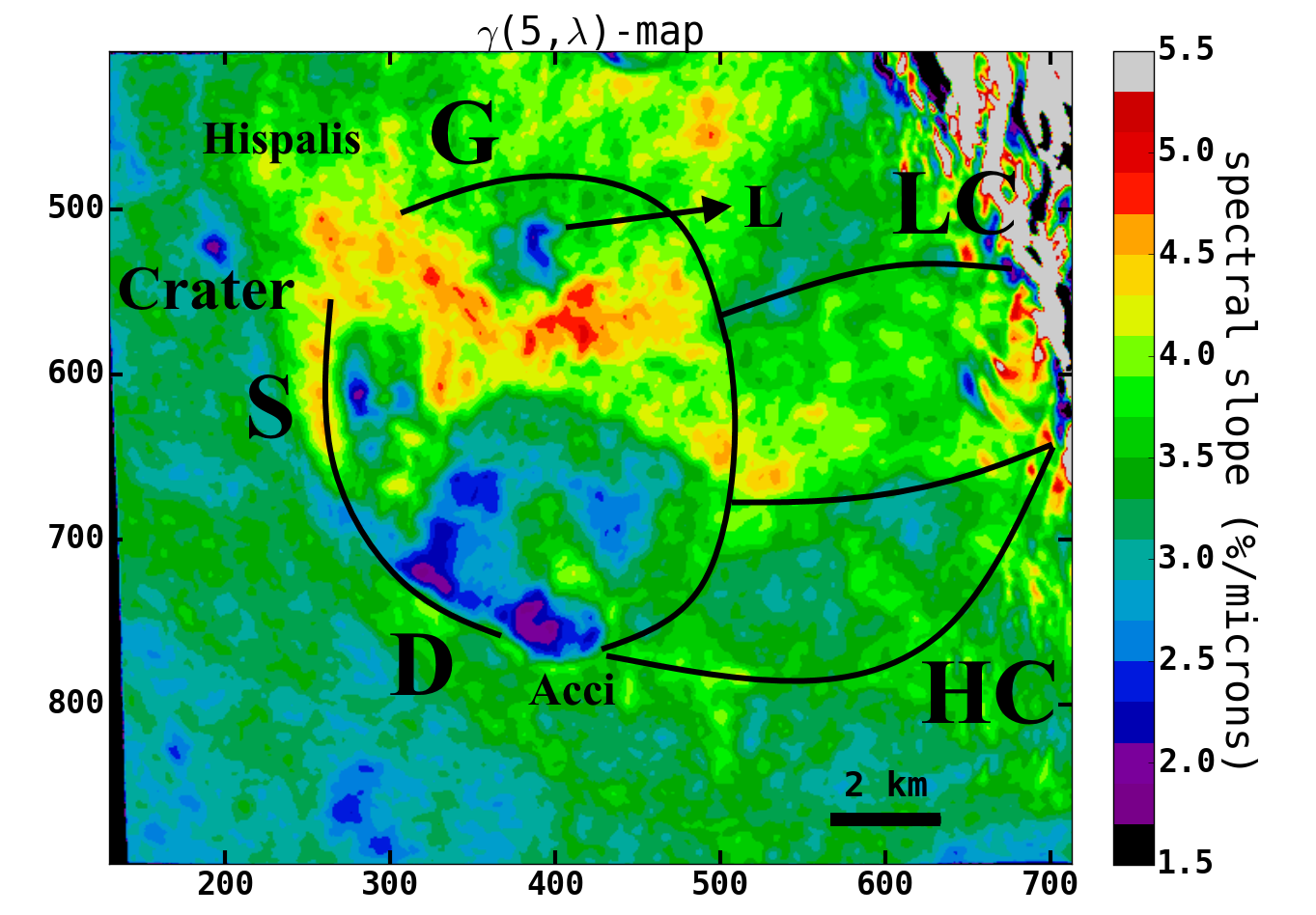}\protect
\par\end{centering}

}\subfloat[]{\protect\begin{centering}
\protect\includegraphics[scale=0.25]{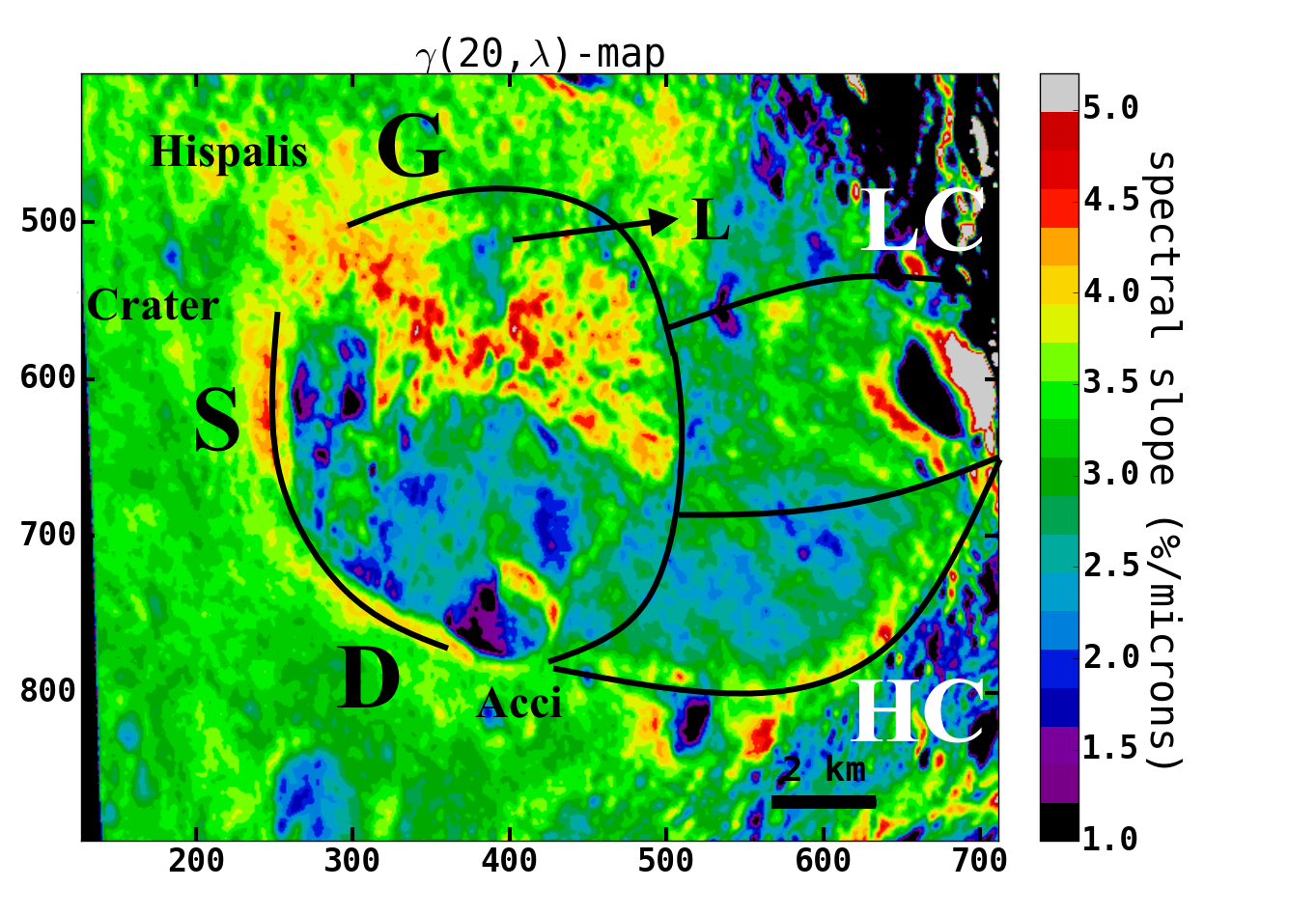}\protect
\par\end{centering}

}
\par\end{centering}

\begin{centering}
\subfloat[]{\protect\begin{centering}
\protect\includegraphics[scale=0.3]{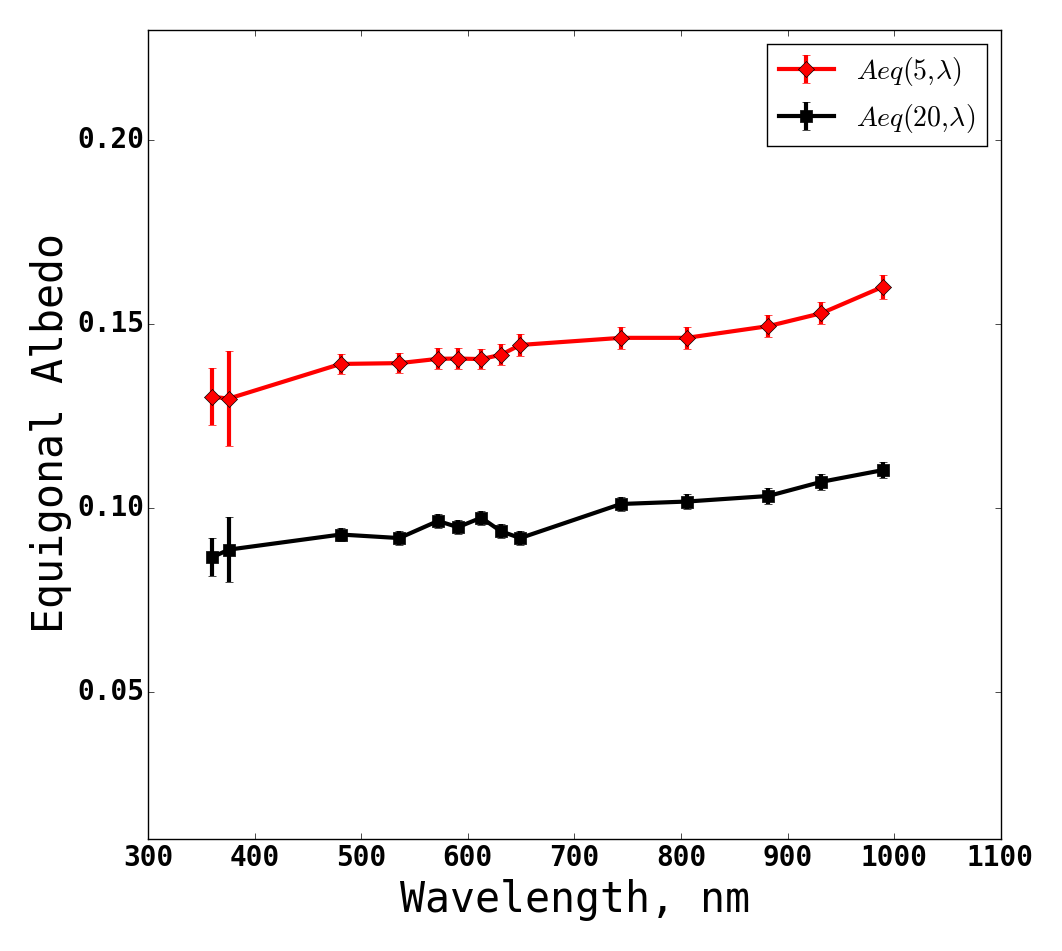}\protect
\par\end{centering}

}
\par\end{centering}

\begin{centering}
\protect\caption{\label{fig:spectral-slope} Projected spatial map of spectral slope
into NAC\_2010-07-10T15.40.39.231Z\_ID30\_1251276000\_F82 image. (a)
Spectral slope $\gamma(\alpha)$ ($\%\cdot\mu m^{-1}$) at phase angle
of $5^{\circ}$ and (b) of $20^{\circ}$. (c) The global $\bar{A}_{eq}(5^{\circ},\lambda)$
and $\bar{A}_{eq}(20^{\circ},\lambda)$ spectrophotometry with OSIRIS
estimated errors. (G) Gallicum. (D) Danuvius. (S) Sarnus. (L) Landslide.
(LC) Low Corduba. (HC) High Corduba.}

\par\end{centering}

\end{figure*}

\par\end{center}

\section{Hapke-parameters resolved map}

\subsection{\textmd{Hapke Model}}

The Hapke model applied in this work is the latest version by Hapke
\citeyearpar{2008Icar..195..918H,9781139025683}. We included some
proposed corrections (Helfenstein \& Shepard, \citeyear{2011Icar..215...83H})
and criticisms \citep{2012JQSRT.113.2431S} in respect to the treatement
of the CBOE. The model attempts to analytically describe many regimes
on the photometric curve of a particulate media based on the most
recent theoretical and laboratory results. The model expression for
the bi-directional reflectance is:

\begin{equation}
\begin{array}{c}
r_{(\mu_{0},\mu,\alpha)}=K\frac{w}{4\pi}\left(\frac{\mu_{0e}}{\mu_{e}+\mu_{0e}}\right)[(1+B_{SH(\alpha)})P_{(\alpha)}\\
+(1+B_{CB(\mu_{0},\mu,\alpha)})(H_{(\mu_{0}/K,w)}H_{(\mu/K,w)}-1)]S_{(\mu_{0},\mu,\alpha)}
\end{array}
\end{equation}
\label{eq:hapke}

\noindent $\mu_{0e}$ and $\mu_{e}$ are the effectives cosine of
incidence and emission angle, involving the topographic correction
of the facet. All Hapke explicit parameters are described in the Table
\ref{tab:hapke-description}. As follows, we summarize the mathematical
framework for all function in expression \ref{eq:hapke}:
\begin{lyxlist}{00.00.0000}
\item [{$P_{(\alpha)}$}] The single-particle phase function is modeled
by the Henyey-Greenstein (H-G) functions \citep{2002Icar..157..523H}.
The functions are single-term or double-term, depending on the kind
of grain is being represented. For double-term function, the asymmetric
factor is given as $g_{sca}=-b\cdot c$.
\item [{$B_{SH}(\alpha,Bs_{0},h_{s})$}] The SHOE function (Section 8,
Hapke, 1993). The half-width at half-maximum (HWHM) of the SHOE is
given by $\Delta\alpha_{SHOE}=2h_{s}$. On our formalism, we explicitly
take $Bs_{0}\propto\frac{S(0)}{wP(0)}$ , as function of the specular
component $S(0)$ (Hapke, 1993). Therefore $Bs_{0}=nS(0)$ , where
$n$ is a unknown proportional component. We observe that normalizing
$Bs_{0}$ give more realistic values, i.e., under unity. Finally,
the porosity factor $K$ and $h_{s}$ are constrained through the
approximative formula $K=1.069+2.109h_{s}+0.577h_{s}^{2}-0.062h_{2}^{2}$
\citep{2011Icar..215...83H}.
\item [{$B_{CB}(\mu_{0},\mu,\alpha,Bc_{0},h_{c})$}] The original CBOE
function by \citet{1988JPhys..49...77A} (equation 18, \citealp{2012JQSRT.113.2431S}).
The HWHM of the CBOE is given by $\Delta\alpha_{CBOE}=0.36h_{c}$.
The CBOE mechanism arises from the multiple-scattering, thus we incorporated
the proposition of Helfenstein \& Shepard \citeyearpar{2011Icar..215...83H}
to make this function multiplying only the multiple-scattering term
in equation \ref{eq:hapke}.
\item [{$H(x,w)$}] The second-order approximate Ambartsumian-Chandrasekhar
function describes the multiple-scattering contribution (equation
13, \citealp{2002Icar..157..523H}).
\item [{$S(\mu_{0},\mu,\alpha)$}] The shadowing function describes the
hindering of brightness due to shadows produced by random irregularities
on the medium. The formulation is long, thus we invite the readers
to consult it in Hapke \citeyearpar{1981JGR....86.4571H,1993tres.book.....H}.
\end{lyxlist}
\begin{center}
\begin{table}
\protect\caption{\label{tab:hapke-description} Characteristics of the Hapke
Parameters.}

\centering{}%
\begin{tabular}{c>{\centering}p{8cm}>{\centering}p{1.6cm}}
{\footnotesize{}parameter} & {\footnotesize{}Description} & {\footnotesize{}Boundaries}\tabularnewline
\hline 
{\footnotesize{}$w$} & {\footnotesize{}Particle Single-Scattering Albedo (SSA).} & {\scriptsize{}$\left\{ 0.05,0.5\right\} $}\tabularnewline
\hline 
{\footnotesize{}$b$} & {\footnotesize{}Coefficients of bi-lobe Heyney-Greenstein function.
$b$ models the wideness of the back and forward lobes.$c$ represents
the partitioned contribution of each. Both coefficients are correlated
by $c=3.29\exp(-17.4b^{2})-0.908$ (Hapke, 2012).} & {\scriptsize{}$\left\{ 0.1,0.9\right\} $}\tabularnewline
\hline 
{\footnotesize{}$g_{sca}$} & {\footnotesize{}Asymmetric factor. Coefficient of the mono-lobe Heyney-Greenstein
function. The average cosine of emergence angle of the particle phase
function.} & {\scriptsize{}$\left\{ -0.9,0.9\right\} $}\tabularnewline
\hline 
{\footnotesize{}$K$} & {\footnotesize{}Porosity factor $K=\ln(1-1.209\phi^{2/3})/1.209\phi^{2/3}$,
$\phi$ is the filling factor. It is an addition introduced by Hapke
(2008) corresponding to the porosity role in the regolith light scattering. } & {\scriptsize{}$\left\{ 1.0,1.6\right\} $}\tabularnewline
\hline 
{\footnotesize{}$Bs_{0}$,$h_{s}$} & {\footnotesize{}Amplitude and angular width of the SHOE. $h_{s}$
relates with micro-roughness and porosity. $Bs_{0}$is a function
of the specular component of the particle scattering function.} & {\scriptsize{}$\left\{ 0.0,1.0\right\} $,}{\scriptsize \par}

{\scriptsize{}$\left\{ 0.0,0.52\right\} $}\tabularnewline
\hline 
{\footnotesize{}$Bc_{0}$,$h_{c}$} & {\footnotesize{}Amplitude and angular width of the CBOE. $h_{c}=\frac{\lambda}{2\pi l}$,
thus $h_{c}$ is inversely proportional to the mean photon path $l$.
$Bc_{0}$ is somehow connect to the particle scattering matrix.} & {\scriptsize{}$\left\{ 0.0,0.6\right\} $,}{\scriptsize \par}

{\scriptsize{}$\left\{ 0.0,0.2\right\} $}\tabularnewline
\hline 
{\footnotesize{}$\bar{\theta}$} & {\footnotesize{}Average macroscopic roughness slope of sub-pixel scale.} & {\scriptsize{}$\left\{ 1^{\circ},60^{\circ}\right\} $}\tabularnewline
\hline 
\end{tabular}
\end{table}

\par\end{center}

The Hapke model is a combination of nonlinear functions with a total
of four to eight free parameters where its application depends on
the bidirectional reflectance sampling, reflectance precision and
illumination coverage of the photometric data. The opposition effect
terms $B_{SH}$ and $B_{CB}$ are generally applied only when reflectance
is measured for $\alpha\lesssim5^{\circ}$, awhile the shadowing function
$S$ is introduced when $i\gtrsim30^{\circ}$ and $\alpha\gtrsim60^{\circ}$.
However, an outcome of inverse problems of nonlinear models is the
lack of unique solutions for a given set of measurements. Usually,
some groups of solutions can be discarded based on previous assumptions
about the nature of the system. In the case of the Hapke model, the
entangling of its parameters, mainly the $w$, $g_{sca}$ and $\bar{\theta}$,
have been widely criticized for a few decades \citep{1989aste.conf..557H}.
Nevertheless \citet{2013JGRE..118..534F}, \citet{2015Icar..253..271F}
and \citet{2015Icar..260...73S} have recently applied a Bayesian
inversion to study the probabilistic density function of each Hapke
parameter and their degeneracy. Their study has found that $w$, $b$,
$c$ and $\bar{\theta}$ follow a unimodal distribution around the
solution and the parameters can be described with a mean and standard
deviation when uncertainties are lower than about 5\%.  \citet{2015Icar..253..271F}
state that a phase angles $>90^{\circ}$ and phase angle coverage
$>50^{\circ}$ are required to constrain the degeneracy, which is,
in fact, satisfied by the OSIRIS data in Baetica region. Since it
is not the scope of the work to test once more the validity of the
Hapke parameters, we have taken \citeauthor{2015Icar..253..271F}
results into consideration when developing our inversion procedure.

\subsection{\textmd{Inversion procedure}}

The Hapke modeling involved independently fitting each facet with
the L-BFGS-B algorithm \citep{BROYDEN01031970,Zhu:1997:ALF:279232.279236}
available through the Scipy package \citep{Scipy}. The L-BFGS-B is
a minimization method that allows setting boundaries to the variables
and uses up to the second derivative to converge to the closest minima.
Table \ref{tab:hapke-description} presents the boundaries used for
the minimization. First, a global modeling is undertake to the binned
data (section 3), as the same fashion applied in \citet{2015arXiv150506888F}.
The global Hapke parameters is used as initial input for each facet.
Then, we fit the facets through computational multiprocessing and
we divide them into three groups, for computation speeding up purposes:
\begin{itemize}
\item Those observed on the opposition regime (phase angle $<7^{\circ}$)
and with incidence angles higher enough to cast large shadow at neighboring
particles ($i>30^{\circ}$). For those, we fit the opposition effect
mechanisms ($B_{SH}$ and $B_{CB}$) and the shadowing function $S(\mu_{0},\mu,\alpha)$.
\emph{Free parameters are $w$, $b$ or $g_{sca}$, $B_{s0}$, $h_{s}$,
$B_{c0}$, $h_{c}$ and $\bar{\theta}$.}
\item Those observed on the opposition regime, but with incidence angles
unable to cast large shadows ($i<30^{\circ}$). The $S(\mu_{0},\mu,\alpha)$
does not diverge from unity in this condition. Then, only the $B_{SH}$
and $B_{CB}$ are fit. \emph{Free parameters are $w$, $b$ or $g_{sca}$,
$B_{s0}$, $h_{s}$, $B_{c0}$ and $h_{c}$.}
\item Those not observed at the opposition regime. We fit only $S(\mu_{0},\mu,\alpha)$.
\emph{Free parameters are $w$, $\bar{\theta}$, $b$ or $g_{sca}$.}
\end{itemize}
Due to the coupled nature of the some Hapke parameters, we avoided
estimating errors for each parameters separately. When the associated
errors to the measurements are small, such as the case for OSIRIS
data, the main causes of parametric uncertainties are the undersampling
of some regimes in the phase curve and the entangling itself. Therefore,
we took the average I/F, $i$ , $e$ and $\alpha$ angles for each
image to simulate the average conditions of a fitting facet. Then,
we monitored the convergence of the error-weighted RMS, first derivative
and parameters, using the same initial input. After a specific turning
point, where the first derivative becomes the smallest and the minimization
procedure scans the minima, we retrieve all the associated Hapke parameters
and calculated their standard deviation in respect to the final solution.

In the next step, we performed three independent tests with intent
to verify the most suitable particle phase function and the role of
CBOE. The tests are:
\begin{lyxlist}{00.00.0000}
\item [{H1}] Hapke model with CBOE and bi-lobe H-G function.
\item [{H2}] Hapke model with CBOE and single-lobe H-G function.
\item [{H3}] Hapke model without CBOE and with single-lobe H-G function.
\end{lyxlist}
\begin{center}
\begin{table*}
\protect\caption{\label{tab:hapke-tests}Hapke average parameters for three tests on
Baetica and Etruria regions at NAC F82+22 (649.5 nm). A global Hapke
(1993) test is present for comparison with the values given by the
updated model.}

\begin{centering}
\begin{tabular}{ccccc}
\hline 
 & {\footnotesize{}H1} & {\footnotesize{}H2} & {\footnotesize{}H3} & {\footnotesize{}Hapke (1993)}\tabularnewline
\hline 
\hline 
{\footnotesize{}CBOE} & {\footnotesize{}Yes} & {\footnotesize{}Yes} & {\footnotesize{}No} & {\footnotesize{}No}\tabularnewline
{\footnotesize{}HG} & {\footnotesize{}bi-lobe} & {\footnotesize{}single-lobe} & {\footnotesize{}single-lobe} & {\footnotesize{}single-lobe}\tabularnewline
{\footnotesize{}$A_{0}$} & {\footnotesize{}$0.204\pm0.005$} & {\footnotesize{}$0.205\pm0.005$} & {\footnotesize{}$0.186\pm0.005$} & {\footnotesize{}$0.196\pm0.005$}\tabularnewline
{\footnotesize{}$w$} & {\footnotesize{}$0.241\pm0.02$} & {\footnotesize{}$0.181\pm0.005$} & {\footnotesize{}$0.182\pm0.005$} & {\footnotesize{}$0.238\pm0.005$}\tabularnewline
{\footnotesize{}$g_{sca}$} & {\footnotesize{}$-0.276\pm0.006$} & {\footnotesize{}$-0.342\pm0.003$} & {\footnotesize{}$-0.343\pm0.006$} & {\footnotesize{}$-0.271\pm0.001$}\tabularnewline
{\footnotesize{}$b$,$c$} & {\footnotesize{}$-0.191\pm0.003$,$1.45\pm0.02$} & {\footnotesize{}-} & {\footnotesize{}-} & {\footnotesize{}-}\tabularnewline
{\footnotesize{}$Bs_{0}$} & {\footnotesize{}$0.872\pm0.02$} & {\footnotesize{}$0.824\pm0.002$} & {\footnotesize{}$0.829\pm0.0025$} & {\footnotesize{}$1.69\pm0.006$}\tabularnewline
{\footnotesize{}$h_{s}$} & {\footnotesize{}$0.086\pm0.004$} & {\footnotesize{}$0.040\pm0.0007$} & {\footnotesize{}$0.039\pm0.00065$} & {\footnotesize{}$0.047\pm0.0002$}\tabularnewline
{\footnotesize{}$Bc_{0}$} & {\footnotesize{}$0.396\pm0.002$} & {\footnotesize{}$0.072\pm0.033$} & {\footnotesize{}-} & {\footnotesize{}-}\tabularnewline
{\footnotesize{}$h_{c}$} & {\footnotesize{}$0.038\pm0.003$} & {\footnotesize{}$0.060\pm0.039$} & {\footnotesize{}-} & {\footnotesize{}-}\tabularnewline
{\footnotesize{}$\bar{\theta}$} & {\footnotesize{}$13.5^{\circ}\pm4.6^{\circ}$} & {\footnotesize{}$11.45^{\circ}\pm3^{\circ}$} & {\footnotesize{}$11.45^{\circ}\pm1^{\circ}$} & {\footnotesize{}$29.2\pm1^{\circ}$}\tabularnewline
{\footnotesize{}$1-\phi$} & {\footnotesize{}$0.788\pm0.003$} & {\footnotesize{}$0.85\pm0.002$} & {\footnotesize{}$0.86\pm0.002$} & {\footnotesize{}-}\tabularnewline
\hline 
$RMS$ &  &  &  & {\footnotesize{}$0.0346$}\tabularnewline
\hline 
\end{tabular}
\par\end{centering}

\end{table*}

\par\end{center}

The CBOE and SHOE functions have similar behavior, both represent
a small peak close to small phase angles. Generally, the CBOE can
be solely confirm at the observation of a sharp negative branch at
small phase angles \citep{2002Icar..159..396S} in the polarimetric
curves or the measurement of dependence of HWHM with wavelength \citep{1988JPhys..49...77A}.
To test that, a initial input for $Bc_{0}$ and $h_{c}$ are set both
to 0.05 ($\Delta\alpha_{CBOE}=1^{\circ}$). We expect no large divergence
from initial condition unless there is any odd opposition peak not
possibly fit solely by the SHOE function. The Table \ref{tab:hapke-tests}
presents the global solutions for each test. At first, we observed
how oddly the $Bc_{0}$ behaves, reaching almost 0.40 at H1 and barely
diverging at H2. $h_{c}$, otherwise, varies 63\%. Since the multiple-scattering
term, for $w=0.2$ and $K=1.2$, contributes for just about 15\% in
the opposition, the factor of 50\% to which the $B_{c0}$ modifies
the multiple-scattering contribution may not be significant in respect
to the other effects, causing the indeterminacy of the parameter.

Although, the addition $B_{CB}(\mu_{0},\mu,\alpha,Bc_{0},h_{c})$
of improves the estimation of the geometric albedo, rising the value
to 9\% when comparing H2 to H3, placing it closer to the thermal geometric
albedo \citep{2012P&SS...66..192O}. Inspecting at the residue across
the images at Figure \ref{fig:hapke-residue}, however, we observe
no large variations at smaller phase angles. H2 and H3 delivers almost
the same percent of residue across the images, and it is overall better
than H1. Therefore, the $B_{SH}$ function describes well the opposition
effect without any necessary addition of another function. Therefore,
aside the geometric albedo, without any noticeably improvement in
the residue comparing H3 over H2, CBOE may still considered ambiguous
under Lutetia opposition effect at NAC F82+22. Nonetheless, we still
choose H2 to represent the Hapke parameters maps produced in the following
subsection. The CBOE parameters may still help interpreting the regolith
properties.

\begin{center}
\begin{figure*}[t]
\begin{centering}
\includegraphics[scale=0.35]{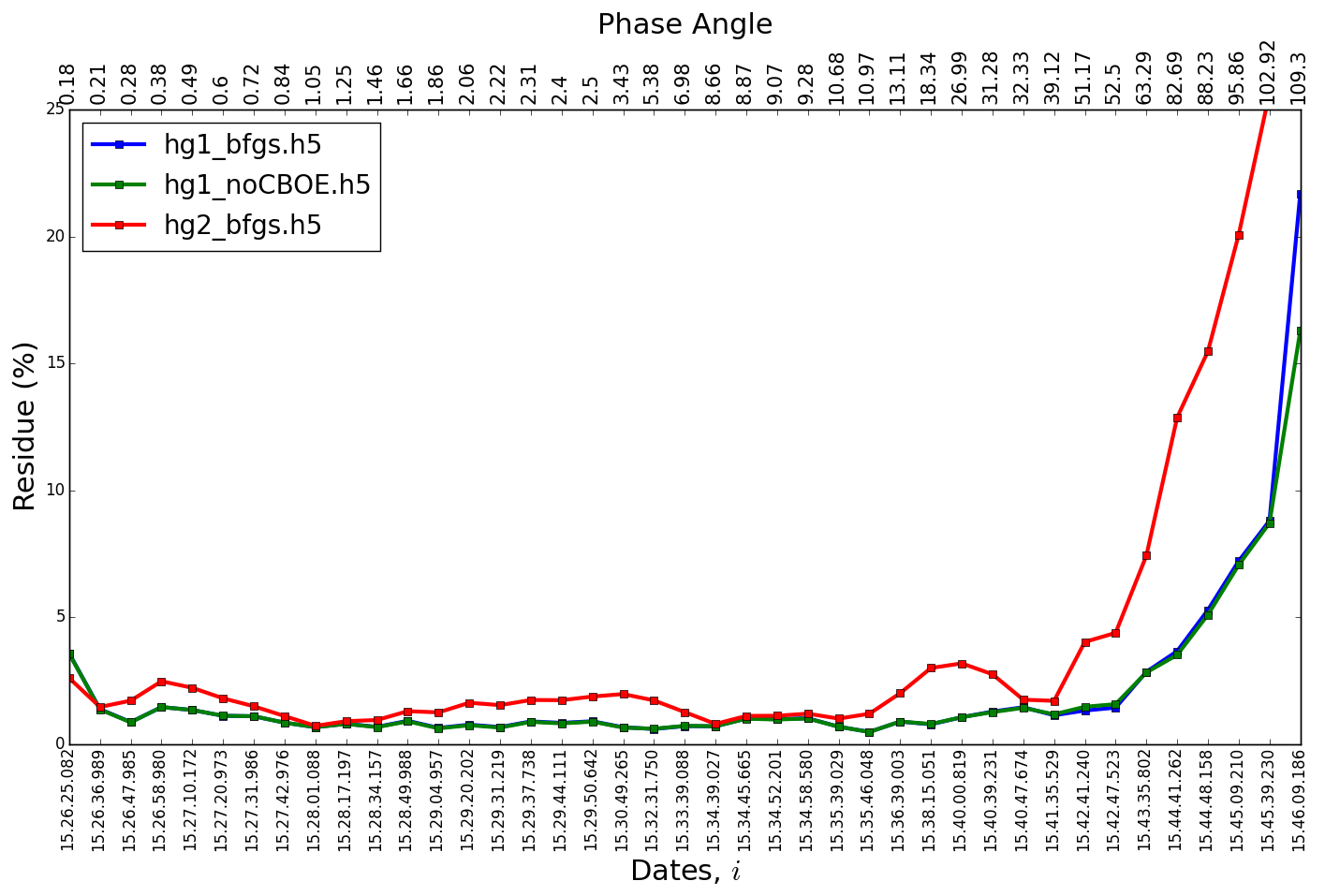}
\par\end{centering}

\protect\caption{\label{fig:hapke-residue} The normalized RMS between the modeled
and measured radiance factor of each image $i$. The residue increases
as the surface becomes darker due to the extreme phase angle and as
shadows overcome great part of the observed facets.}

\end{figure*}

\par\end{center}

\subsection{\textmd{Analysis for the NAC F82+22 filter}}

\begin{center}
\begin{figure*}[ph]
\begin{centering}
\includegraphics[scale=0.1]{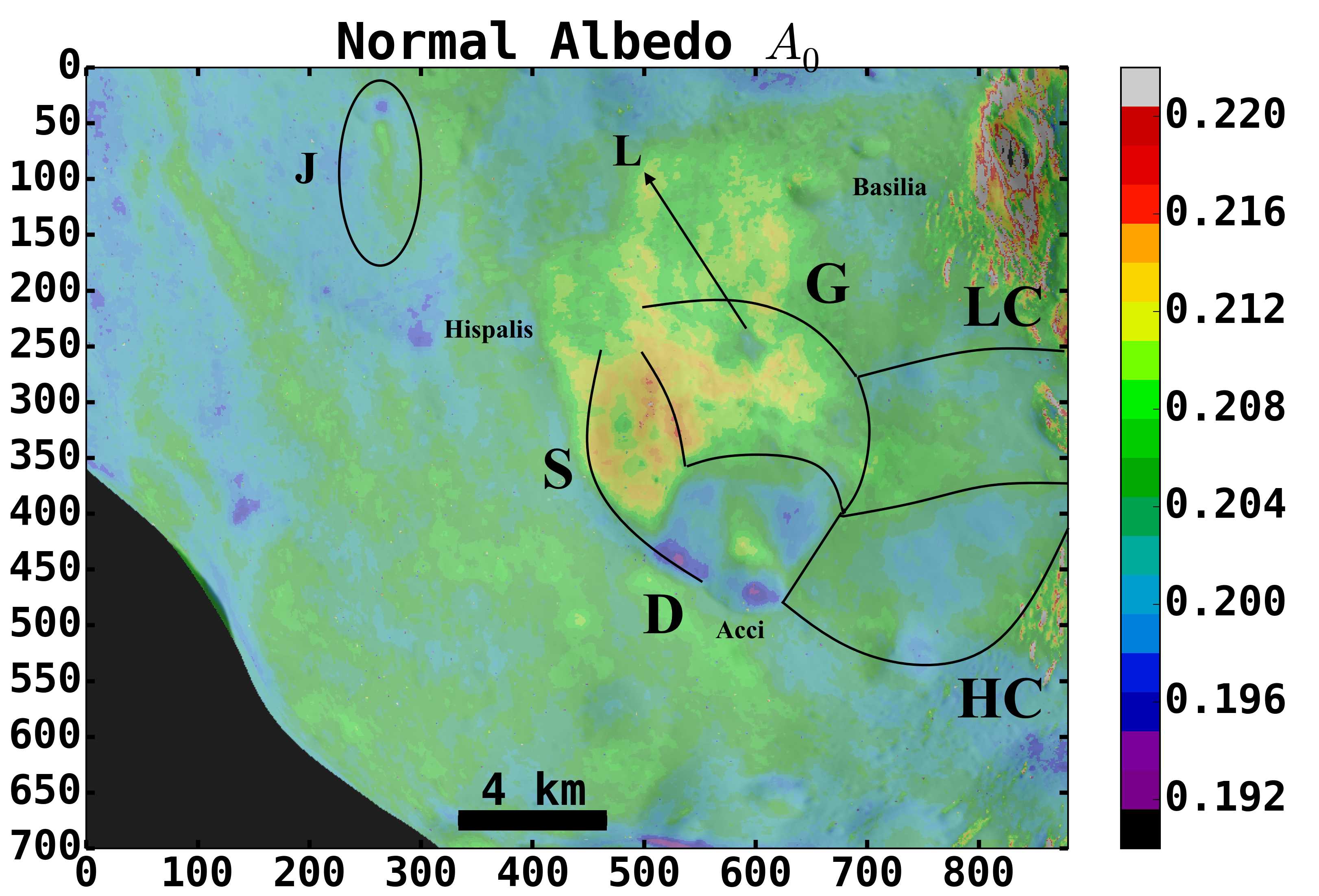}\includegraphics[scale=0.1]{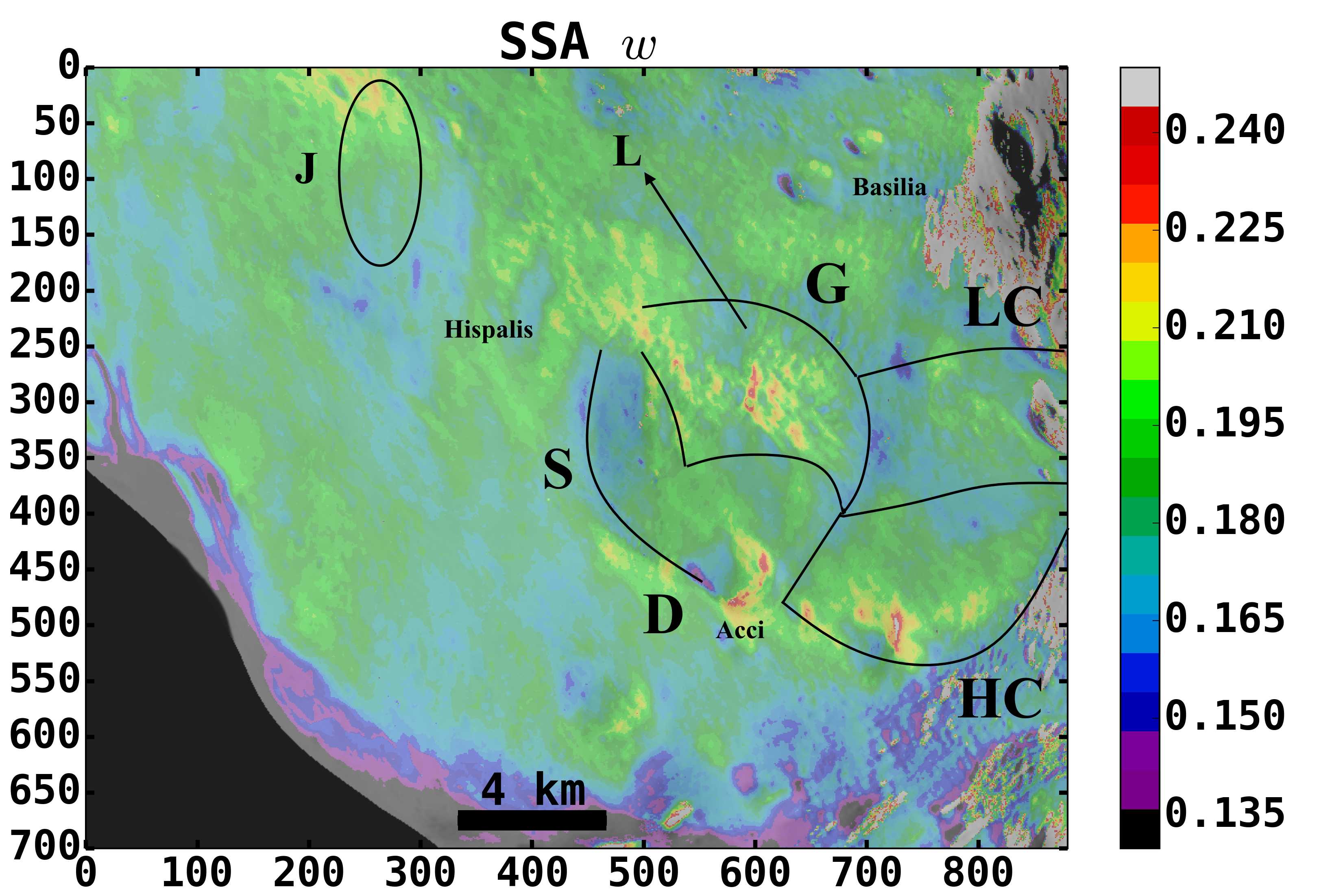}
\par\end{centering}

\begin{centering}
\includegraphics[scale=0.1]{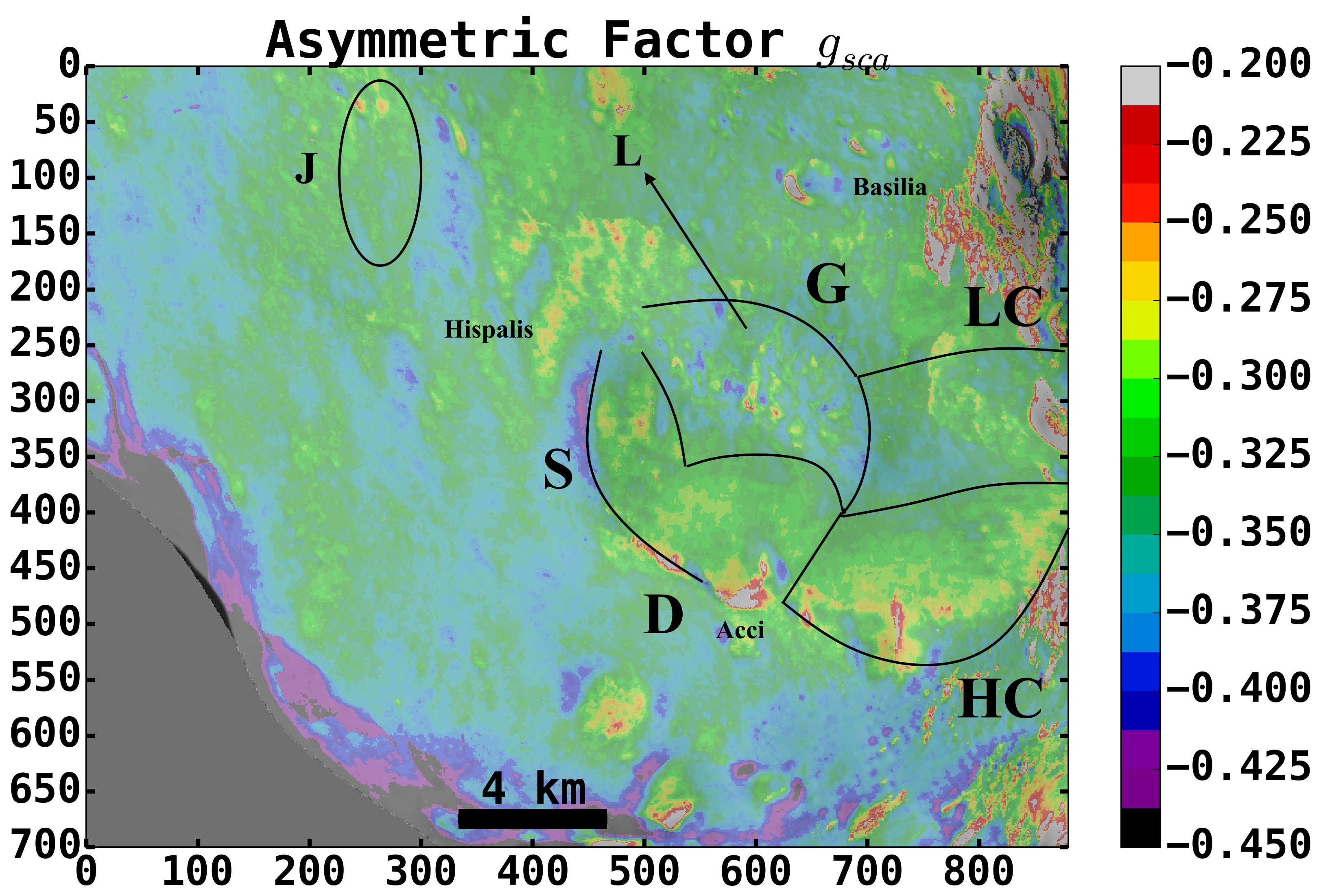}\includegraphics[scale=0.1]{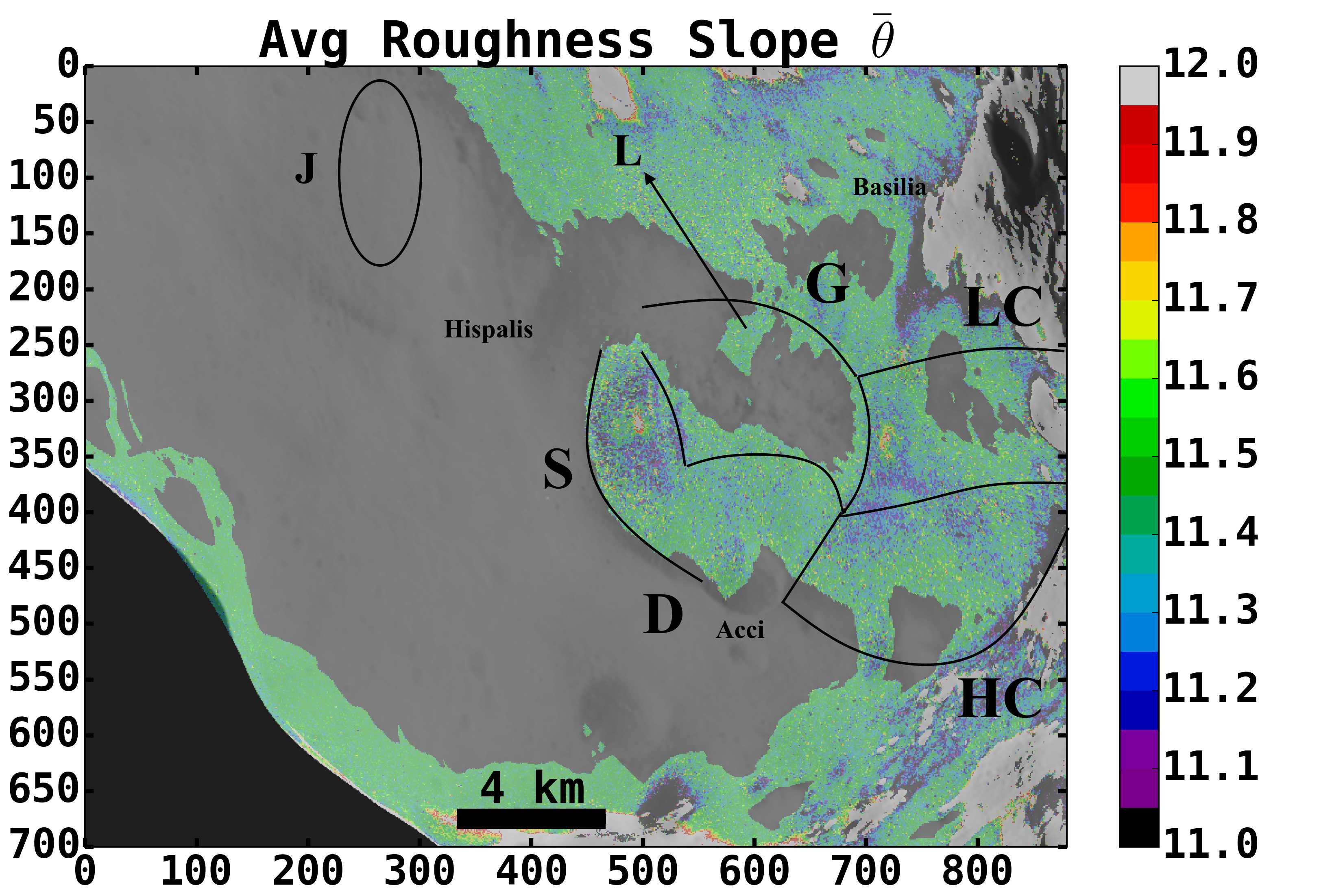}
\par\end{centering}

\begin{centering}
\includegraphics[scale=0.1]{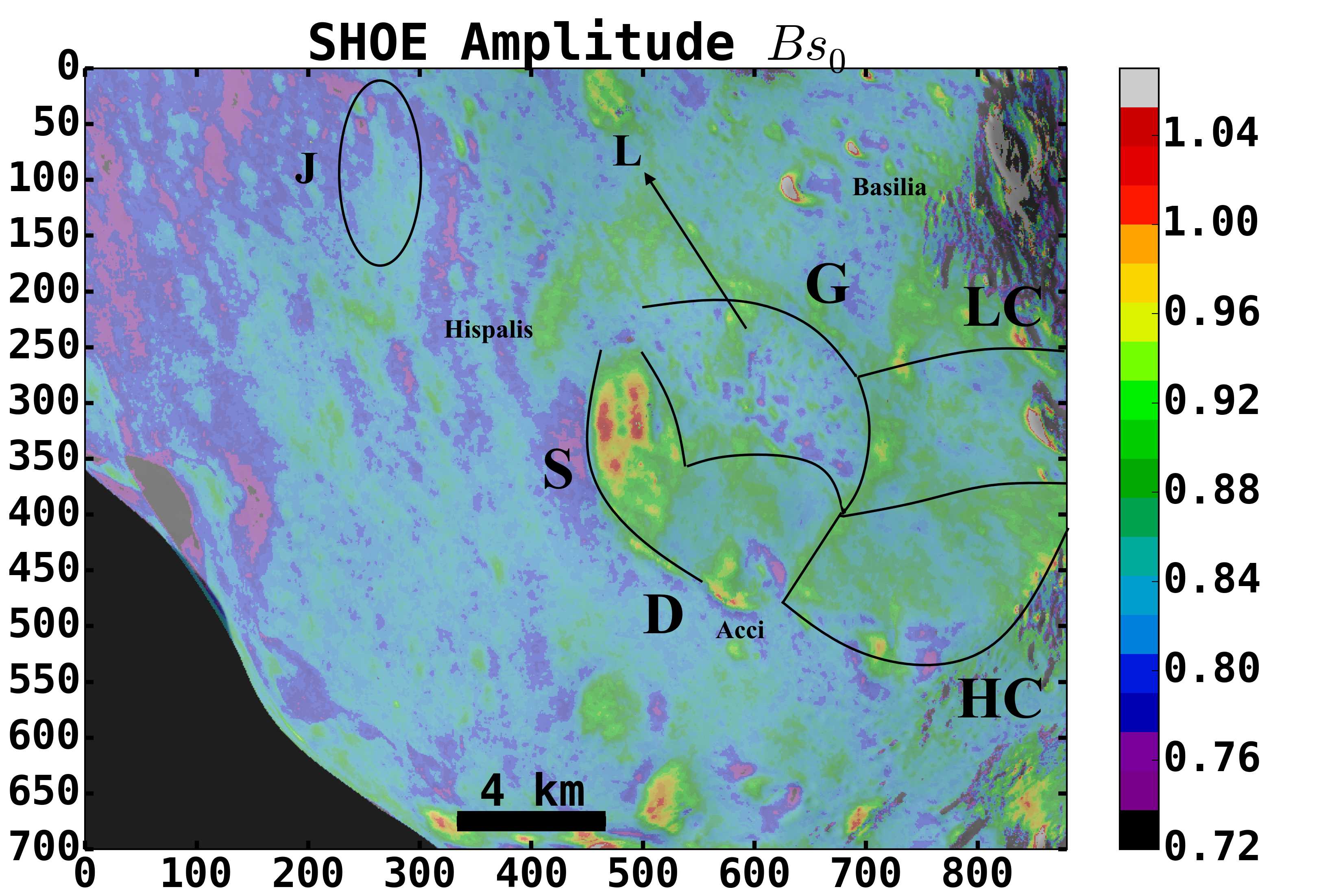}\includegraphics[scale=0.1]{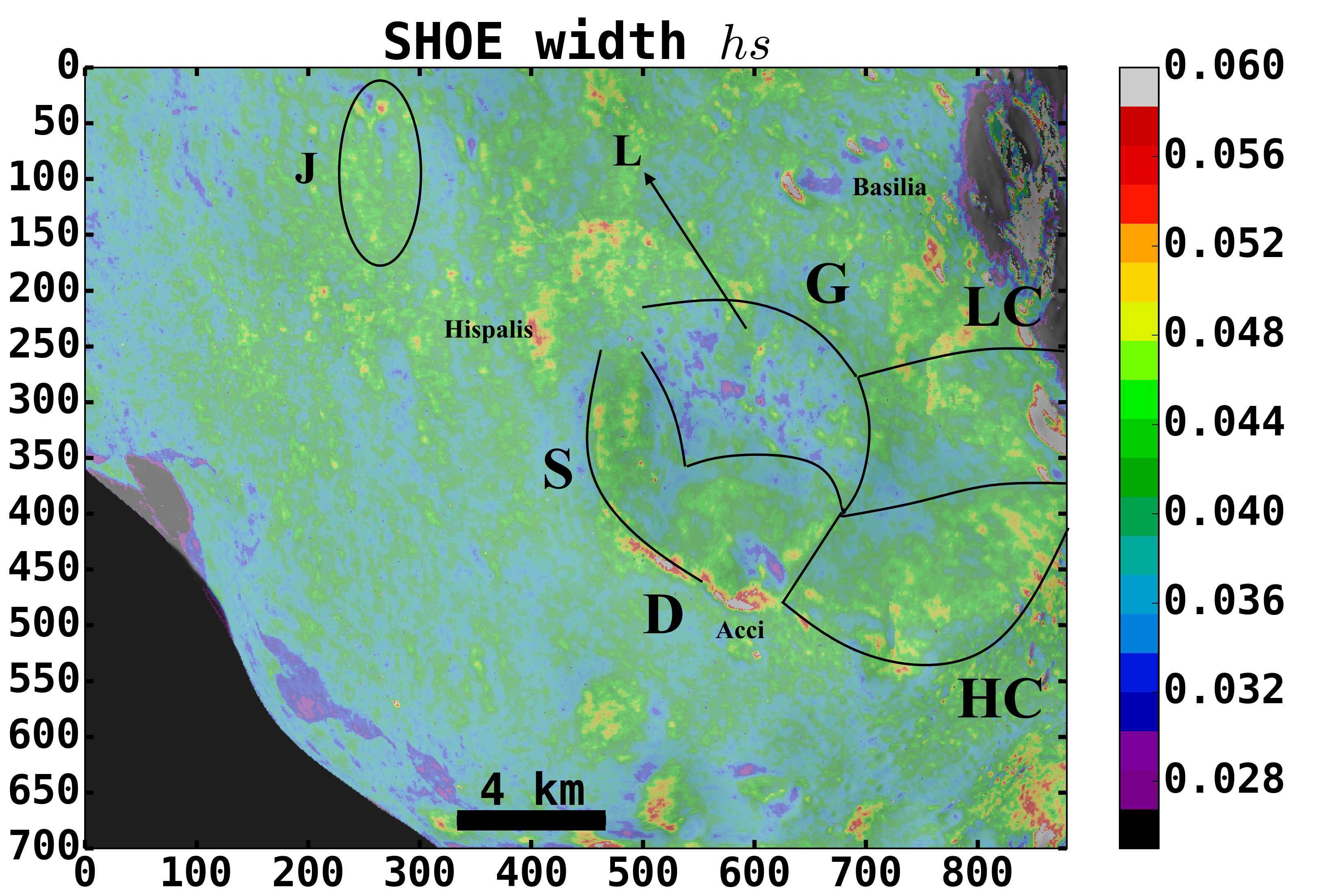}
\par\end{centering}

\protect\caption{\label{fig:hapke-parameters} Resolved Hapke parameters for 650 nm
(NAC F82+F22) projected to NAC\_2010-07-10T15.40.39.231Z\_ID30\_1251276000\_F82
image. The projections are labeled according the parameter. Narbonensis
region at the right corner returns the most problematic values and
highest residue due to the phase angle coverage starting at $34.3^{\circ}$
and the high incidence angles of $80^{\circ}$. (G) Gallicum. (D)
Danuvius. (LH) Low Corduba. (HC) High Corduba. (S) Sarnus. (L) Landslide.
(J) Ejecta.}

\end{figure*}

\par\end{center}

The projected Hapke parameters, normal albedo and normalized RMS into
NAC\_2010-07-10T15.40.39.231Z\_ID30\_1251276000\_F82 image are shown
in the Figure \ref{fig:hapke-parameters}. Some limb, border and crater
effects should be ignored due to the method failing to describe extreme
illumination and multiple-scattering on strong convex topographies.
We observe that Gades and Corduba contains the strongest features
in the Baetica region. The average normal albedo of Sarnus Labes,
Gallicum Labes are and High Corduba is about 8\% higher than the global
average of the region. Danuvius and High Corduba, on the other side,
have a normal albedo of about 2-3\% smaller. The differences are also
reproduced on the Hapke parameters, where the four regions have their
own particular behavior. 

Gallicum Labes and Low Corduba, the brighter and rougher NPCC talus
on the raw images \citep{2012P&SS...66...96T}, are observed on low
incidence angle condition, out of the Hapke shadowing function $S$
regime. Its parameters present a broader CBOE and a sharper SHOE,
a 33\%-higher SSA and a $Bs_{0}$ value closer to average and Danuvius
Labes ($\bar{B}s_{0}=0.81\pm0.03$), which can be interpreted as presence
of less opaque, smoother and fine-grained regolith, corresponding
to a increase in the internal multiple scattering. This kind of regolith
distribution could be related to background soil properties, awhile
large settled boulders have insignificant photometric contribution
in the scattering. On the spatial distribution of the parameters and
on the normal albedo map, we can observe a alignment of bright stripped
material coming down from Hispalis to Gallicum and Sarnus, likely
related to material once in the bottom of NPCC.

Measured on incidence angles of about $70^{\circ}$, allowing us to
estimate $\bar{\theta}$, the Sarnus Labes has also a different behavior
from Gallicum and Danuvius. It presents a broad CBOE and SHOE, high
$Bc_{0}$ and $Bs_{0}$amplitudes, 11\%-darker single-scattering albedo
and 8\%-higher normal albedo indicates a intermediary regolith where
fine-grained particles co-exist with other large opaque particles.
Visually, Sarnus is quite similar to Gallicum, which might denote
that a broad SHOE could be an effect of the intermediary incidence
angle, that turns internal shadows more expressive on this lightning
condition. The $\bar{\theta}\approx12^{\circ}$, nonetheless, is close
to the average of Baetica region, which does not present any wide
variation.

\begin{center}
\begin{figure*}[p]
\begin{centering}
\includegraphics[scale=0.1]{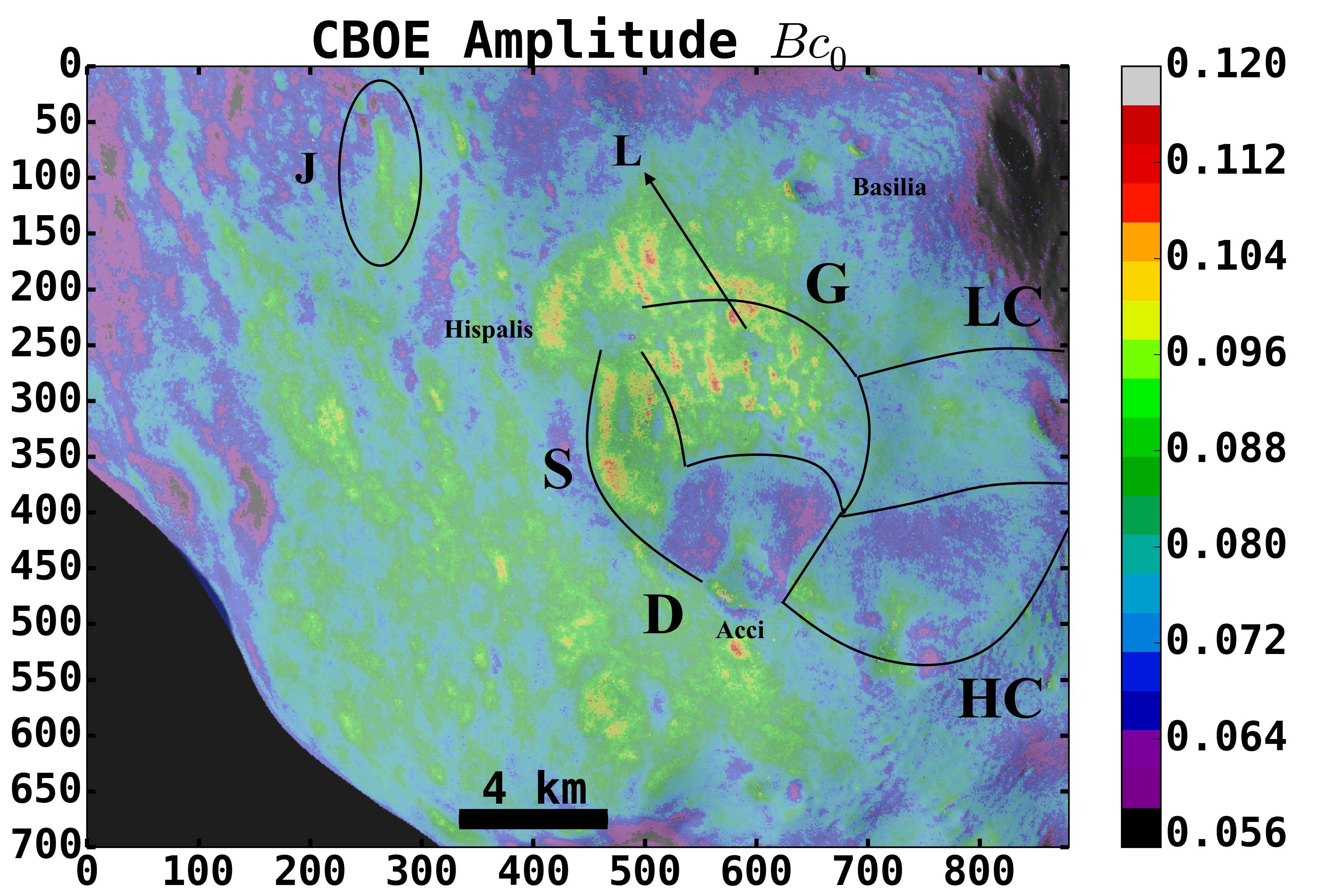}\includegraphics[scale=0.1]{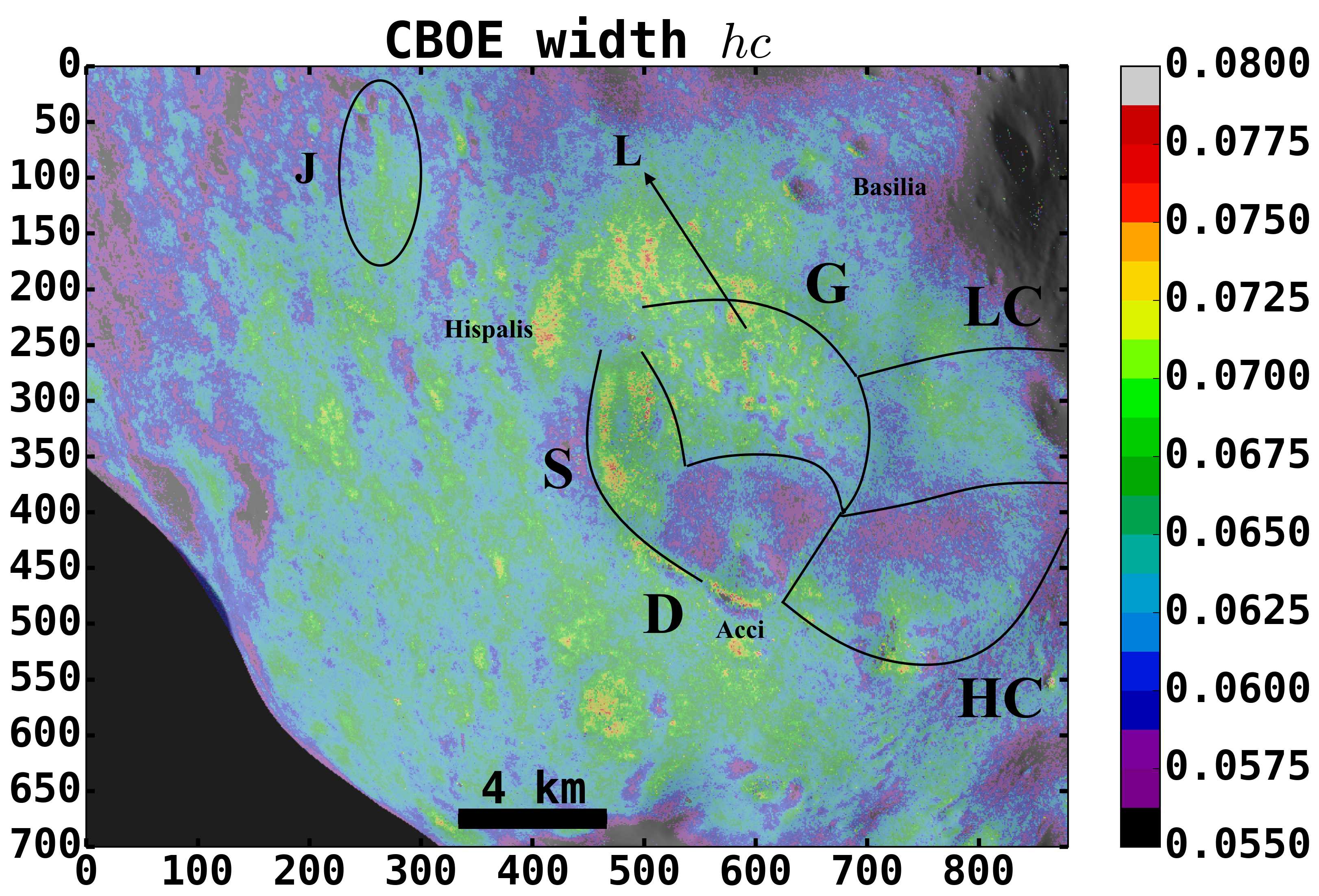}
\par\end{centering}

\begin{centering}
\includegraphics[scale=0.1]{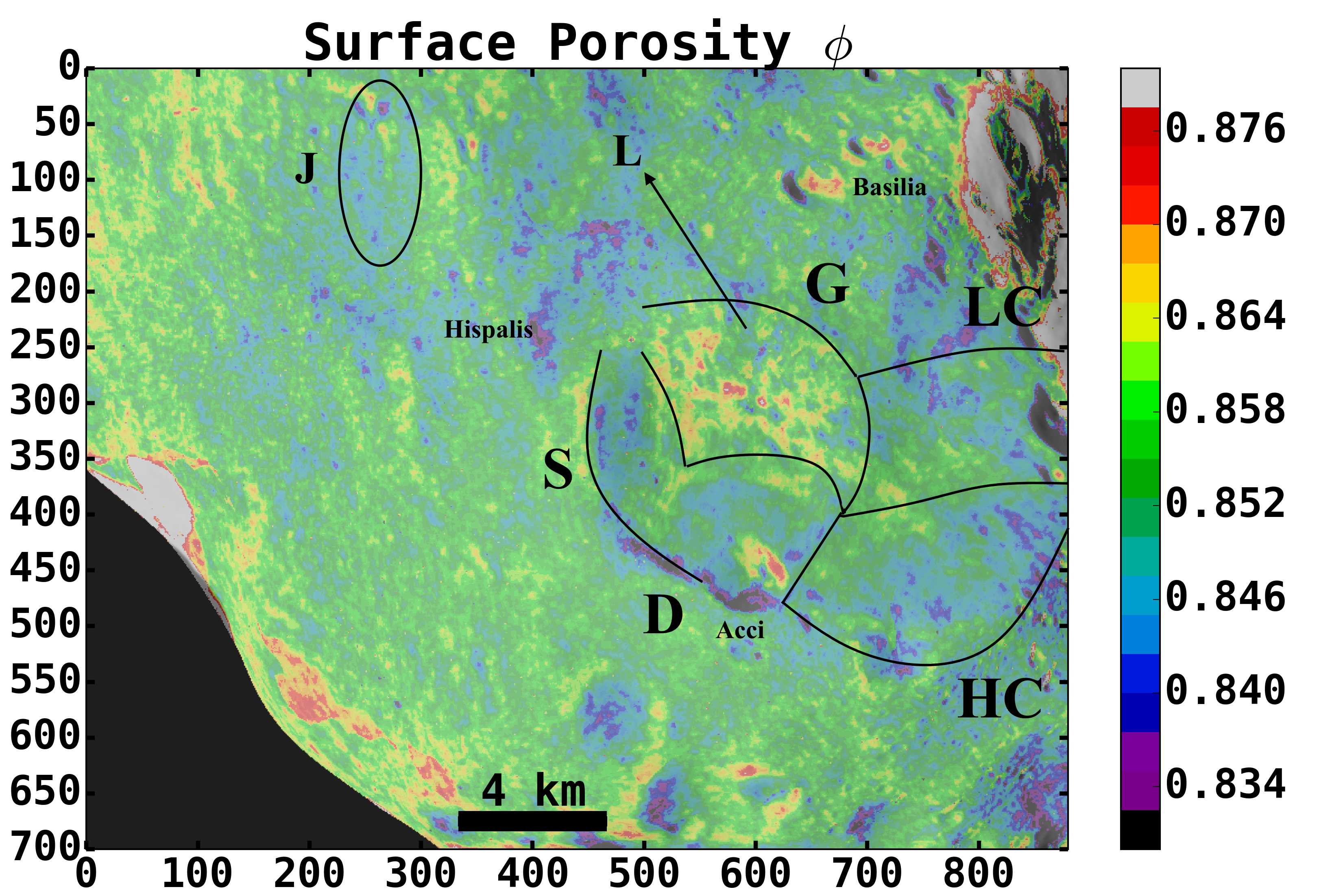}\includegraphics[scale=0.1]{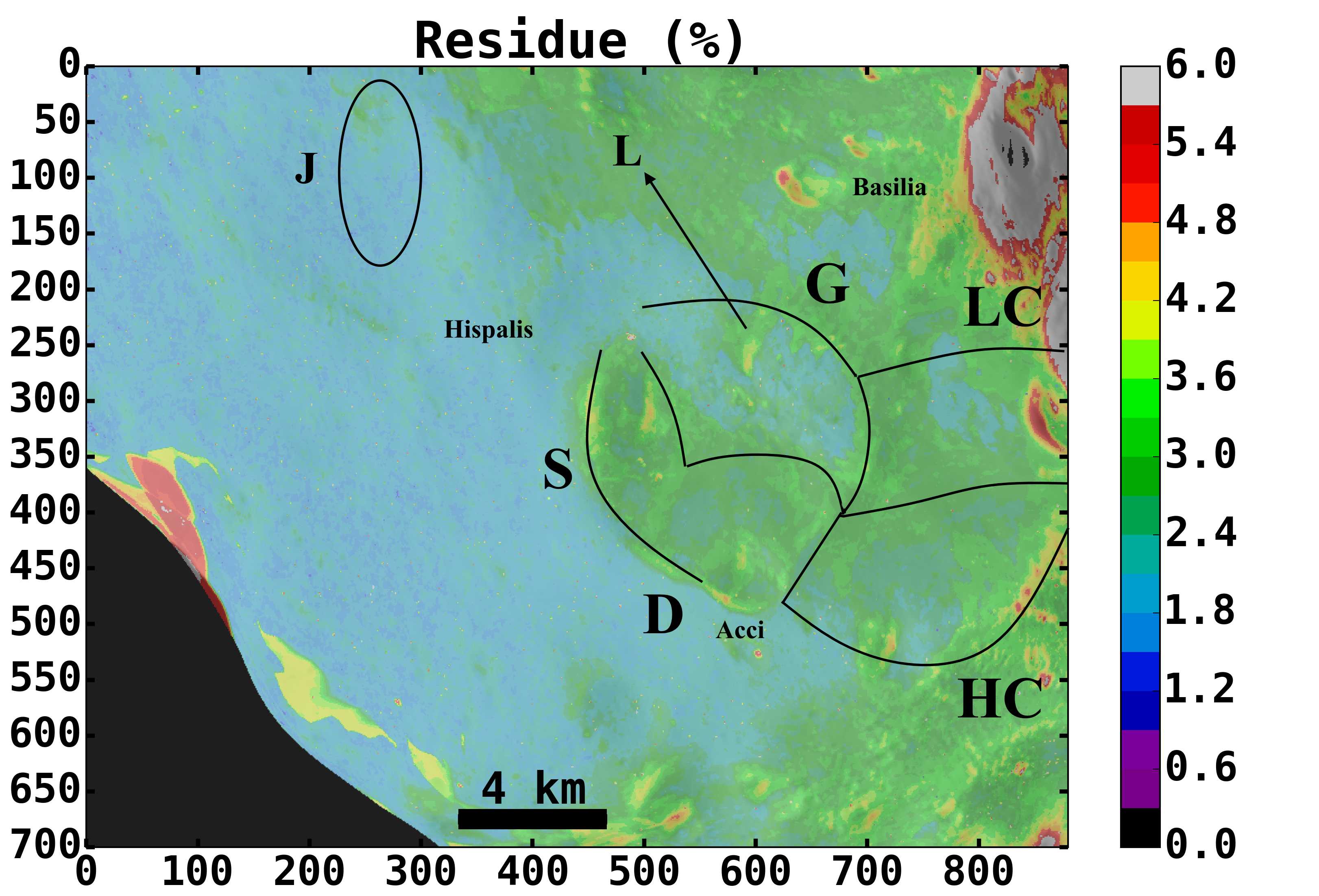}
\par\end{centering}

\begin{centering}
Cont. Figure 4.2.
\par\end{centering}

\end{figure*}

\par\end{center}

Danuvius Labes and High Corduba are observed at intermediary incidence
angles of $45^{\circ}$. Morphologically, Danuvius and High Corduba
are smother, absent of larger boulder and presents a wavy-pattern.
The wall has a slightly darker normal albedo crossed by a 3\%-brighter
strip. This strip is probably connected to the Acci crater and it
is also reproduced in the SSA and CBOE parameter maps. The most discrepant
parameter in the area is the $g_{sca}$, which is 11\% smaller than
average.

 Globally for Baetica and part of Etruria, the SHOE mechanism seems
completely prevalent on the North pole. The single-scattering albedo
is close to the normal albedo, resulting of just about 15\% of contribution
by incoherent multiple-scattering in the opposition. The $\bar{\theta}$
is low. The $g_{sca}$ value represent back-scatter grains, corresponding
to rough-surfaced grains \citep{1995Icar..113..134M} or a regolith
with significant inter-grain porosity \citep{2011Icar..215..526S}.
Therefore, we can state that Lutetia is largely composed of irregular,
opaque and larger-than-wavelength grains under a smooth and porous
superficial distribution. 

We complement that Etruria region, despite being out of the scope
of our work, is partially modeled and presents a interesting bright
strip across this region (Structure J). The bright structure starts
out of a crater and landslides into a bigger depression. The structure
J is discernible in the $A_{0}$, $w$, $Bc_{0}$, $h_{c}$ and porosity
maps. It is possibly the ejecta material of a recent crater, which
have not yet suffered a spectral attenuation due to space weathering.
The material shows similar characteristics to Gallicum Labes and surroundings,
but much less brighter.

\subsection{\textmd{Analysis for the WAC F13 and WAC F17 filters}}

\begin{center}
\begin{table}
\protect\caption{\label{tab:Hapke-wac}Hapke parameters of WAC F13 (375.6 nm) and WAC
F17 (630 nm) for Baetica and Etruria region. H2 test applied to both
filters.}

\begin{centering}
\begin{tabular}{ccc}
\hline 
WAC & F13 & F17\tabularnewline
\hline 
\hline 
{\footnotesize{}CBOE} & {\footnotesize{}Yes} & Yes\tabularnewline
{\footnotesize{}HG} & {\footnotesize{}single-lobe} & {\footnotesize{}single-lobe}\tabularnewline
{\footnotesize{}$A_{0}$} & {\footnotesize{}$0.184\pm0.02$} & {\footnotesize{}$0.194\pm0.01$}\tabularnewline
{\footnotesize{}$w$} & {\footnotesize{}$0.165\pm0.02$} & {\footnotesize{}$0.179\pm0.01$}\tabularnewline
{\footnotesize{}$g_{sca}$} & {\footnotesize{}$-0.346\pm0.026$} & {\footnotesize{}$-0.349\pm0.025$}\tabularnewline
{\footnotesize{}$b$,$c$} & {\footnotesize{}-} & {\footnotesize{}-}\tabularnewline
{\footnotesize{}$Bs_{0}$} & {\footnotesize{}$0.765\pm0.015$} & {\footnotesize{}$0.788\pm0.013$}\tabularnewline
{\footnotesize{}$h_{s}$} & {\footnotesize{}$0.034\pm0.002$} & {\footnotesize{}$0.035\pm0.0013$}\tabularnewline
{\footnotesize{}$Bc_{0}$} & {\footnotesize{}$0.0645\pm0.003$} & {\footnotesize{}$0.062\pm0.017$}\tabularnewline
{\footnotesize{}$h_{c}$} & {\footnotesize{}$0.055\pm0.014$} & {\footnotesize{}$0.057\pm0.07$}\tabularnewline
{\footnotesize{}$\bar{\theta}$} & {\footnotesize{}$16.9^{\circ}\pm5^{\circ}$} & {\footnotesize{}$14.04^{\circ}\pm5^{\circ}$}\tabularnewline
{\footnotesize{}$1-\phi$} & {\footnotesize{}$0.864\pm0.005$} & {\footnotesize{}$0.863\pm0.005$}\tabularnewline
\hline 
 &  & \tabularnewline
\hline 
\end{tabular}
\par\end{centering}

\end{table}

\par\end{center}

Lutetia was observed with WAC F13 (375 nm) and WAC F17 (630 nm) 30
and 39 times, respectively (Table \ref{tab:OSIRIS-images}). The images
were obtained mainly close to the opposition. The oversampling in
this regime allows us a good estimation the Normal Albedo and the
opposition parameters. Therefore, we applied the same Hapke analysis
undertaken to NAC F82+22, using the conditions of test H2. The Table
\ref{tab:Hapke-wac} presents the average Hapke parameters. The Hapke
parameter maps are similar to NAC F82+F22. We find a significant spectral
variation of the single-scattering albedo and the Normal Albedo (Figure
\ref{fig:normal-albedo}). We fit a line for each corresponding pixel
in the two filters plus the NAC F82+F22, in the same fashion described
in section 3.2. The $w$ spectral slope map and the $A_{0}$ spectral
slope map (or $\gamma(0^{\circ})$-map) presents an average of $\bar{\gamma}_{w}(0^{\circ})=6\pm1.15\%\cdot\mu m^{-1}$
and of $\bar{\gamma}(0^{\circ})=5.8\pm1.35\%\cdot\mu m^{-1}$, respectively. 

The $\gamma(0^{\circ})$-map shows the spectral dichotomy of the Gallicum
Labes-Low Corduba and Danuvius labes-High Corduba, as previously stated.
We measure a phase bluing of $\frac{\bar{\gamma}(20^{\circ})-\bar{\gamma}(0^{\circ})}{20^{\circ}-0^{\circ}}=-0.145\pm0.08\%\cdot\mu m^{-1}deg^{-1}$.
However, we are dealing with just 3 filters instead of the 14 used
in the $\gamma(5,\lambda)$ and $\gamma(20,\lambda)$ maps. If we
consider using only the same NAC F82+22, WAC F13 and WAC F17 filters
for $\gamma(20,\lambda)$ map, hence the average is then decreased
to $\bar{\gamma}(20^{\circ})=2.1\pm1.92\%\cdot\mu m^{-1}$. It represents
a higher phase bluing ($-0.185\pm0.13\%\cdot\mu m^{-1}deg^{-1}$),
but uncertainties are also increased due to the decrease of filter
number, making the phase bluing less statistically significant. 

A weak phase bluing, no higher than 3\%, has been reported by \citet{2014Icar..239..201S}
for some basalt powders between the opposition and mid-phase angles
($15^{\circ}-30^{\circ}$). The effect is more evident at 1000 nm/800
nm ratio for pressed samples and probably connected to the SHOE, since
the particles were fully opaque during the experiment. Furthermore,
among asteroid surfaces, phase bluing have also been observed for
44 Nysa by \citet{2009Icar..201..655R} in the U-B and V-R colors.
In the U-B, a color opposition effect is observed, indicating a difference
in the OE angular width in respect to the wavelength.

The $w$ spectral slope map shows some distortion due to the border
and small craters, but Sarnus Labes bears a redder slope of about
25\% than average. The $w$ and $A_{0}$ spectral slopes seems decoupled,
which means that not all spectral wavelength dependence observe in
the region is due to single-scattering. Therefore, we expect to observe
the dependence of $h_{c}$ with the wavelength, which it is considered
an evidence in favor of the presence of CBOE. However, the spectral
slope for $h_{c}(\lambda)$ carries a large degree of uncertainty
($16\pm13\%\cdot\mu m^{-1}$) and most of the spectro-morphological
variations  are confined up to $\pm5.6\%\cdot\mu m^{-1}$.

\begin{center}
\begin{figure}
\begin{centering}
\subfloat[]{\protect\begin{centering}
\protect\includegraphics[scale=0.35]{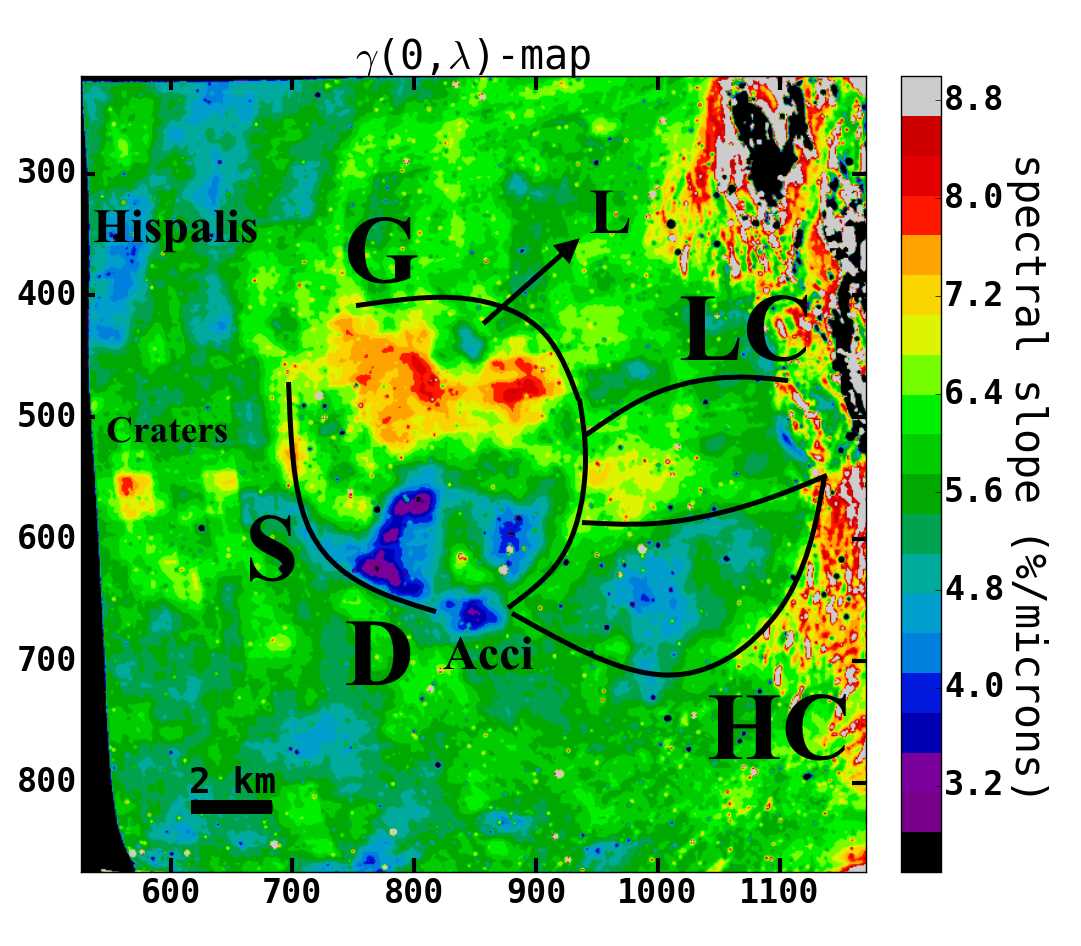}\protect
\par\end{centering}

}
\par\end{centering}

\begin{centering}
\subfloat[]{\protect\begin{centering}
\protect\includegraphics[scale=0.37]{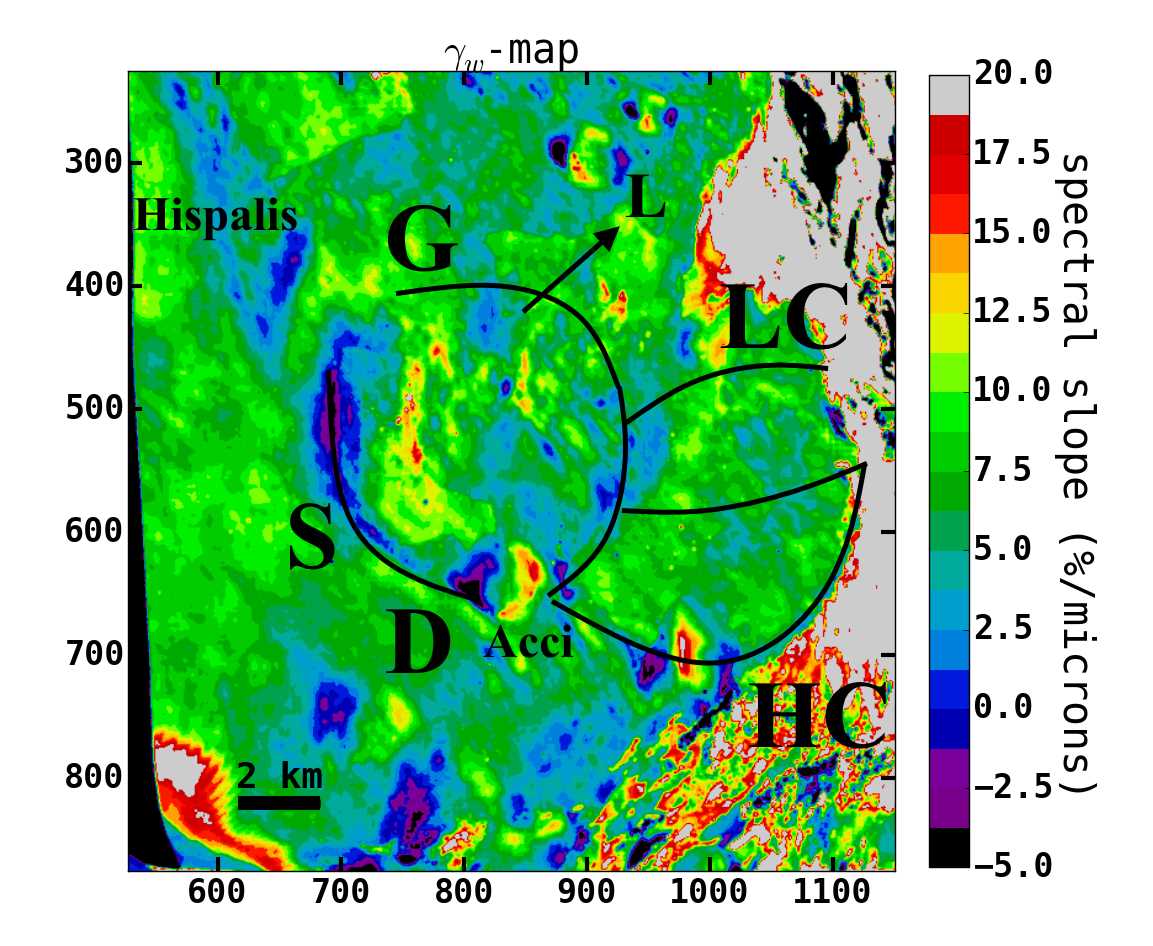}\protect
\par\end{centering}

}
\par\end{centering}

\protect\caption{\label{fig:normal-albedo} Spectrophotometric slope $\gamma$ of (a)
the Normal Albedo and (b) the single-scattering albedo. Projected
into NAC\_2010-07-10T15.40.39.231Z\_ID30\_1251276000\_F82 image. (G)
Gallicum. (D) Danuvius. (S) Sarnus. (HC) High Corduba. (LC) Low Corduba.}

\end{figure}

\par\end{center}

\section{Discussion}

\subsubsection*{The presence of CBOE}

Our results are, in respect to the presence of the CBOE, ambiguous.
The Hapke CBOE parameters do not improve the fitting and the SHOE
only is enough to fit the opposition surge. Although we find an expressive
phase bluing in the $A_{0}$ spectral slope map. \citet{2010A&A...515A..29B}
work on polarimetry of Lutetia was also inconclusive in respect to
the existence of coherent-backscattering mechanism. Belskaya et al.
showed that Lutetia presents one of broadest negative polarization
branch, only (234) Barbara and other four asteroids have wider negative
polarization than Lutetia. When comparing Lutetia polarimetric curve
with laboratory powder of meteorites, only the carbonaceous chondrites,
specially the CV3, show similar minimum of negative branch and inversion
angle. Multiple hypothesis have been evoked to explain this phenomena
(\citealp{1994EM&P...65..201S}, for a review), but the shadow-hiding
and coherent-backscattering seems to prevail as the main mechanisms
\citep{2002Icar..159..396S}. The connection with such dark meteoritic
composition should rule out the contribution of CBOE, since the mechanism
is also connect to grain shape and size. \citet{2002Icar..159..396S}
states that scattering of higher orders still significantly contribute
to negative polarization, even for very dark surfaces. Therefore,
the wide negative branch of Lutetia may still be a product of both
mechanisms.

\subsubsection*{A comparison with the Akimov phase parameters and the Hapke parameters}

We have calculated the Spearman's rank correlation coefficient among
the Akimov phase parameters and Hapke parameters (Figure \ref{fig:akimov-hapke}).
The Spearman's rank measures respectively, from -1 to +1, the degree
of anti-correlation and correlation of two parameters of a sample.
As expected, the $A0_{Akimov}$ and the $A0_{Hapke}$ are strongly
correlated. $m_{Akimov}$ and $w_{1,Akimov}$ describe the phase curve
non-linear slope, where the first is more related to CBOE term and
the latter to SHOE term. We have $w_{2,Akimov}$ linked to parameters
someway related to micro-roughness, mainly $g_{sca}$ and $h_{s}$.
It has been shown that micro-roughness is intrinsically connected
to surface porosity \citep{2007JGRE..112.3001S}, which is a propriety
correlated to $h_{s}$ in \citet{2008Icar..195..918H}. Thereby the
link to porosity/$h_{s}$ corroborates the interpretation given for
$w_{2,Akimov}$ by \citet{2011P&SS...59.1326S}.

\begin{center}
\begin{figure}
\begin{centering}
\includegraphics[scale=0.4]{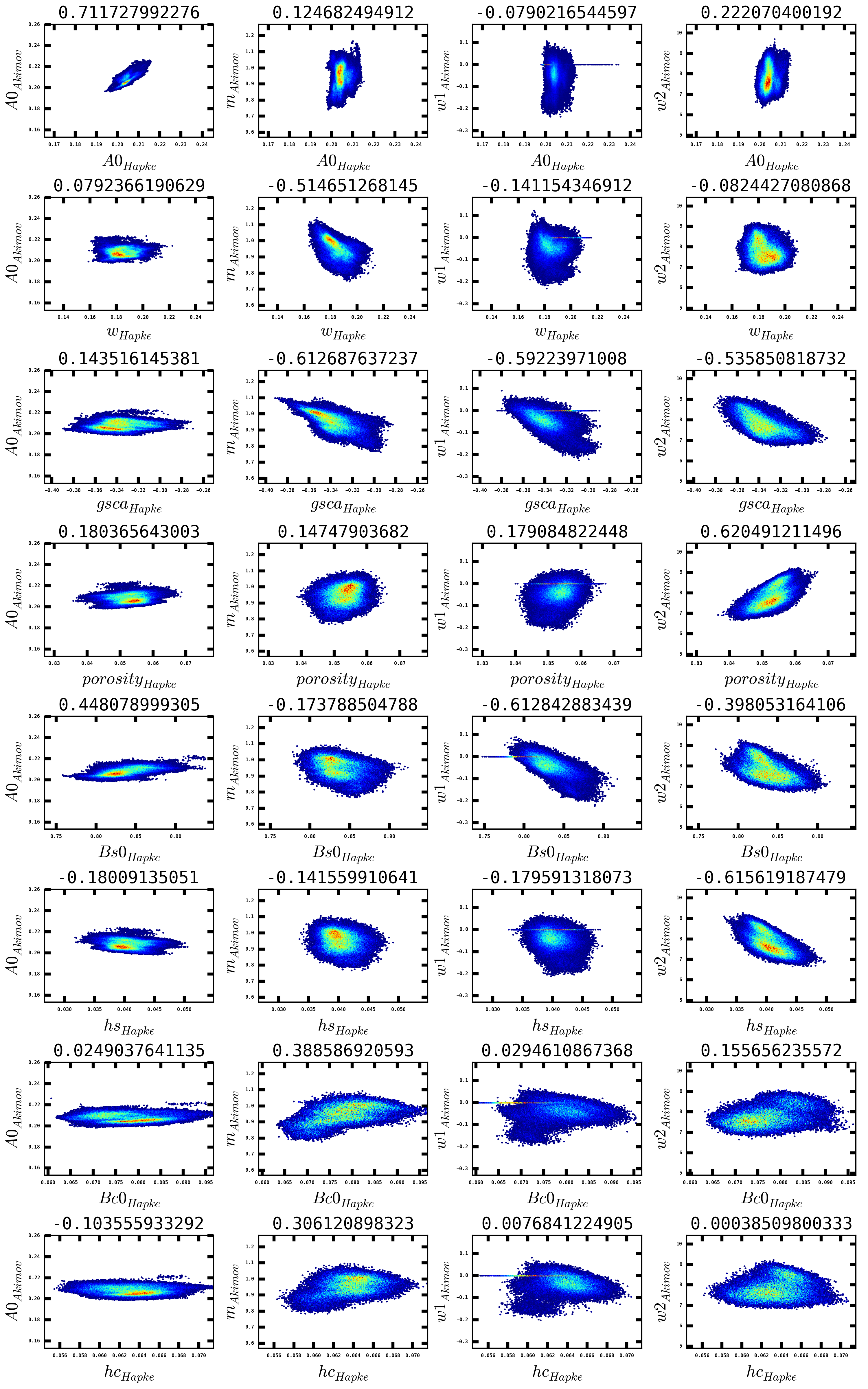}
\par\end{centering}

\protect\caption{\label{fig:akimov-hapke}Correlation of pixels of the projected
map in the NAC\_2010-07-10T15.40.39.231Z\_ID30\_1251276000\_F82 image
for Akimov phase parameters and Hapke parameters in the NAC F82+22
filter. The parameters are calculated for each facet then translated
to pixels according to the methodology described on subsections 3.1
and 4.2. The title on each plot references to the respective Sperman's
rank correlation coefficient.}

\end{figure}

\par\end{center}

\subsubsection*{A comparison with laboratory measurement and Global regolith properties}

The precise estimation of Hapke parameters of (21) Lutetia allow us
compare it to laboratory samples where their Hapke parameter have
been determined. \citet{2007JGRE..112.3001S} and \citet{2011Icar..215..526S}
have tested Hapke model to particulated media of several different
characteristics: compression, grain size, roughness, shape and composition.
In both paper the authors have accounted for the CBOE, but only in
\citet{2011Icar..215..526S} used the same Hapke formalism and accounted
for the porosity correction. \citet{2012Icar..218..364B} have applied
the \citet{1993tres.book.....H} model for reflectance phase curve
of 8 meteoritic samples. The authors did not include the opposition
effect mechanisms, which allows comparisons with just three parameters:
$w$ , $g_{sca}$ and $\bar{\theta}$. While we do not state that
any of the samples are, in fact, what Lutetia is composed of, we expect
they give us some analog information about the major characteristics
of the regolith covering Lutetia in the Baetica region.

None of the laboratory samples of the cited works have matching Hapke
parameters we have derived for Baetica. The meteoritic samples were
compared to our Hapke (1993) parameters, and likewise, none of them
match neither. However, only Cobalt Oxide measured at 700 nm (CO 700)
and Red Iron Oxide measured at 550 nm (FE 550) are the particulated
media with most similar $w$, $h_{s}$, $Bs_{0}$, $\bar{\theta}$
and porosity (their parameters are produced in Table \ref{tab:objects}). 

The CO 700 is described as a dark, flat spectrum and dense powder
of $\sim20\ \mu m$ spheroidal grains. Its SSA, $h_{s}$ and porosity
encloses well Lutetia parameters, considering the errors. $\bar{\theta}$
is defined as $16^{\circ}$ in \citet{2007JGRE..112.3001S} against
$0^{\circ}$in \citet{2011Icar..215..526S}. The farther is closer
to Lutetia's. On the high-resolution optical images, CO presents some
roughness and large grains on its surface, thus $\bar{\theta}=0^{\circ}$
is probably incorrect. On the scanning electron microscope (SEM) images,
the grain are indeed more ellipsoidal and of sharp cut shape.

FE 550 is composed of primarily red-spectrum hematite and it is described
as composed of shape-complex grains of around $4-10$ microns in size.
Its SSA, $h_{s}$, $Bs_{0}$ and $b$ are the closest Hapke parameters
to Lutetia's. The $\bar{\theta}$ is almost the double, but is encompassed
by the average value found on asteroids (Table\ref{tab:objects}).
The high-resolution optical images show an very rough surface, populated
by large grains and micro shadows. The SEM images present intersected
grains of extreme irregular shape.

However, CO 700 show no similar CBOE parameters. The test presents
too high $h_{c}$of corresponding HWHM of $27.5^{\circ}$, such large
value probably interferes on the estimation of the particle phase
function parameters. FE 550, on the other hand, is ambiguous about
the CBOE, the RMS is relatively independent of the presence of the
mechanism.

\subsubsection*{A comparison to Masoumzadeh et al. work}

\citet{2015Icar..257..239M} has recently undertaken a global photometric
analysis of all observed surfaces of (21) Lutetia using Minnaert disk
function and \citet{1993tres.book.....H} model. They have analyzed
the same OSIRIS NAC F22, NAC F82 and WAC F17 images we used to develop
this present work. Their Minnaert analysis has shown an unusual non-linear
behavior of the disk function in (21) Lutetia, meaning that Lutetia
surface becomes slightly more lambertian at larger phase angles. The
$k$ coefficient found in this work for zero phase angle ($k_{H2015}(0)=0.5505\pm0.01$)
match the one found by them ($k_{M2015}(0)=0.526\pm0.002$).

The \citet{1993tres.book.....H} $g_{sca}$, $\bar{\theta}$ and $h_{s}$
parameters obtained in this work and \citeauthor{2015Icar..257..239M}
are in good agreement (Table \ref{tab:objects}), whilst $w$ and
$Bs_{0}$ values diverge for just about 6\% and 8\%, respectively.
These results point to a healthy convergence on both analysis and
applied inversion methodology.

When presenting the albedo maps and phase ratios, \citeauthor{2015Icar..257..239M}
state that no large-scale variegation higher than 7\% aside on NPCC
was detected. Our normal albedo and parameter maps support such statements,
where large variations on (21) Lutetia are contained in Gades and
Corduba craters. However, although no higher than 3\% of the the average
normal albedo, we detect some small-scale variations in Etruria as
the Structure J.

\subsubsection*{A comparison with other small solar system bodies}

We compiled in the Table\ref{tab:objects} the global Hapke parameters
of Small Solar System Bodies modeled after disk-resolved data. To
be able of comparing the global Hapke parameters of Lutetia with other
bodies, we also fit the Hapke (1993) model into data (Table \ref{tab:hapke-tests}).
Lutetia shares similar $w$ , $g_{sca}$, $A_{0}$ and $Bs_{0}$ with
(243) Ida, a S-type asteroid of 31.4 km diameter. $\bar{\theta}$
and $A_{0}$ are also close to (941) Gaspra and (433) Eros, both main-belt
asteroids of S group. The resemblance of Lutetia phase curve to S-type
asteroids has been previously pointed out by \citet{2015Icar..257..239M}.
However, we must emphasize we do not state that Lutetia compositionally
match to S-type asteroids. The parameters may only indicate that those
asteroid share near regolith properties, like grain size distribution,
compression and roughness. 

Most of the bodies seems to have $Bs_{0}$ > 1, which has been interpreted
as necessity to included the CBOE mechanism. None of them present
a completely matching opposition effect characteristics ($Bs_{0}$
and $h_{s}$). The $Bs_{0}$ of Lutetia is the second higher after
(253) Mathilde among the Main-belt asteroids.

\begin{center}
\begin{sidewaystable}
\protect\caption{\label{tab:objects}Global Hapke parameters of other Small Solar System
Bodies. Values in parenthesis are fixed. }

\begin{centering}
\begin{tabular}{ccccccccccccc}
\hline 
{\scriptsize{}Object} & {\scriptsize{}Type} & {\scriptsize{}$w$} & {\scriptsize{}$g_{sca}$} & {\scriptsize{}$\bar{\theta}$} & {\scriptsize{}$Bs_{0}$} & {\scriptsize{}$h_{s}$} & {\scriptsize{}$Bc_{0}$} & {\scriptsize{}$h_{c}$} & {\scriptsize{}$1-\phi$} & {\scriptsize{}$A_{0}$} & {\scriptsize{}$\lambda_{nm}$} & {\scriptsize{}Reference}\tabularnewline
\hline 
\hline 
{\scriptsize{}Lutetia, Baetica} & {\scriptsize{}-} & {\scriptsize{}0.181} & {\scriptsize{}-0.343} & {\scriptsize{}11.45} & {\scriptsize{}0.824} & {\scriptsize{}0.04} & {\scriptsize{}0.072} & {\scriptsize{}0.06} & {\scriptsize{}85} & {\scriptsize{}0.205} & {\scriptsize{}650} & {\scriptsize{}this work}\tabularnewline
\hline 
{\scriptsize{}Lutetia} & {\scriptsize{}M/C} & {\scriptsize{}0.238} & {\scriptsize{}-0.271} & {\scriptsize{}29} & {\scriptsize{}1.69} & {\scriptsize{}0.047} & {\scriptsize{}-} & {\scriptsize{}-} & {\scriptsize{}-} & {\scriptsize{}0.196} & {\scriptsize{}650} & {\scriptsize{}this work (Hapke, 1993)}\tabularnewline
\hline 
{\scriptsize{}Lutetia} & {\scriptsize{}M/C} & {\scriptsize{}0.22} & {\scriptsize{}-0.28} & {\scriptsize{}28} & {\scriptsize{}1.78} & {\scriptsize{}0.05} & {\scriptsize{}-} & {\scriptsize{}-} & {\scriptsize{}-} & {\scriptsize{}-} & {\scriptsize{}650} & {\scriptsize{}Masoumzadeh et al. \citeyearpar{2015Icar..257..239M}}\tabularnewline
{\scriptsize{}Cobalt Oxide} & {\scriptsize{}sample} & {\scriptsize{}0.186} & {\scriptsize{}-0.33} & {\scriptsize{}0/16} & {\scriptsize{}1.00} & {\scriptsize{}0.08} & {\scriptsize{}0.681} & {\scriptsize{}0.7} & {\scriptsize{}83} & {\scriptsize{}-} & {\scriptsize{}700} & {\scriptsize{}Helfenstein \& Shepard \citeyearpar{2011Icar..215..526S}}\tabularnewline
{\scriptsize{}Iron Oxide} & {\scriptsize{}sample} & {\scriptsize{}0.17} & {\scriptsize{}-0.09} & {\scriptsize{}22} & {\scriptsize{}(1.00)} & {\scriptsize{}0.07} & {\scriptsize{}-} & {\scriptsize{}-} & {\scriptsize{}-} & {\scriptsize{}-} & {\scriptsize{}550} & {\scriptsize{}Shepard \& Helfenstein \citeyearpar{2007JGRE..112.3001S}}\tabularnewline
{\scriptsize{}(1) Ceres} & {\scriptsize{}G} & {\scriptsize{}0.07} & {\scriptsize{}-0.4} & {\scriptsize{}44} & {\scriptsize{}1.58} & {\scriptsize{}0.06} & {\scriptsize{}-} & {\scriptsize{}-} & {\scriptsize{}-} & {\scriptsize{}0.088} & {\scriptsize{}555} & {\scriptsize{}Li et al. \citeyearpar{2006Icar..182..143L}}\tabularnewline
{\scriptsize{}(4) Vesta} & {\scriptsize{}V} & {\scriptsize{}0.51} & {\scriptsize{}-0.24} & {\scriptsize{}18} & {\scriptsize{}1.7} & {\scriptsize{}0.07} & {\scriptsize{}-} & {\scriptsize{}-} & {\scriptsize{}-} & {\scriptsize{}0.42} & {\scriptsize{}554} & {\scriptsize{}Li et al. \citeyearpar{2013Icar..226.1252L}}\tabularnewline
{\scriptsize{}(243) Ida} & {\scriptsize{}S} & {\scriptsize{}0.22} & {\scriptsize{}-0.33} & {\scriptsize{}18} & {\scriptsize{}1.53} & {\scriptsize{}0.02} & {\scriptsize{}- } & {\scriptsize{}-} & {\scriptsize{}-} & {\scriptsize{}0.21} & {\scriptsize{}560} & {\scriptsize{}Helfenstein et al. \citeyearpar{1996Icar..120...48H}}\tabularnewline
{\scriptsize{}(253) Mathilde} & {\scriptsize{}C} & {\scriptsize{}0.035} & {\scriptsize{}-0.25} & {\scriptsize{}19} & {\scriptsize{}3.18} & {\scriptsize{}0.074} & {\scriptsize{}-} & {\scriptsize{}-} & {\scriptsize{}-} & {\scriptsize{}0.041} & {\scriptsize{}700} & {\scriptsize{}Clark et al. \citeyearpar{1999Icar..140...53C}}\tabularnewline
{\scriptsize{}(433) Eros} & {\scriptsize{}S} & {\scriptsize{}0.33} & {\scriptsize{}-0.25} & {\scriptsize{}28} & {\scriptsize{}1.4} & {\scriptsize{}0.01} & {\scriptsize{}-} & {\scriptsize{}-} & {\scriptsize{}-} & {\scriptsize{}0.23} & {\scriptsize{}550} & {\scriptsize{}Li et al. \citeyearpar{2004Icar..172..415L}}\tabularnewline
{\scriptsize{}(941) Gaspra} & {\scriptsize{}S} & {\scriptsize{}0.36} & {\scriptsize{}-0.18} & {\scriptsize{}29} & {\scriptsize{}1.63} & {\scriptsize{}0.06} & {\scriptsize{}-} & {\scriptsize{}-} & {\scriptsize{}-} & {\scriptsize{}0.22} & {\scriptsize{}560} & {\scriptsize{}Helfenstein et al. \citeyearpar{1994Icar..107...37H}}\tabularnewline
{\scriptsize{}(2867) Steins} & {\scriptsize{}E} & {\scriptsize{}0.57} & {\scriptsize{}-0.27} & {\scriptsize{}28} & {\scriptsize{}0.6} & {\scriptsize{}0.06} & {\scriptsize{}0.52} & {\scriptsize{}0.0025} & {\scriptsize{}84} & {\scriptsize{}0.39} & {\scriptsize{}630} & {\scriptsize{}Spujth et al. \citeyearpar{2012Icar..221.1101S}}\tabularnewline
{\scriptsize{}(5535) Annefrank} & {\scriptsize{}S} & {\scriptsize{}0.63} & {\scriptsize{}-0.09} & {\scriptsize{}50} & {\scriptsize{}(1.32)} & {\scriptsize{}0.015} & {\scriptsize{}-} & {\scriptsize{}-} & {\scriptsize{}-} & {\scriptsize{}0.28} & {\scriptsize{}550} & {\scriptsize{}Hillier et al. \citeyearpar{2011Icar..211..546H}}\tabularnewline
{\scriptsize{}(25143) Itokawa} & {\scriptsize{}S} & {\scriptsize{}0.42} & {\scriptsize{}-0.35} & {\scriptsize{}26} & {\scriptsize{}0.87} & {\scriptsize{}0.01} & {\scriptsize{}-} & {\scriptsize{}-} & {\scriptsize{}-} & {\scriptsize{}0.33} & {\scriptsize{}1570} & {\scriptsize{}Kitazato et al. \citeyearpar{2008Icar..194..137K}}\tabularnewline
{\scriptsize{}Moon, Highland} & {\scriptsize{}-} & {\scriptsize{}0.38} & {\scriptsize{}-0.09} & {\scriptsize{}24} & {\scriptsize{}1.7} & {\scriptsize{}0.075} & {\scriptsize{}-} & {\scriptsize{}-} & {\scriptsize{}-} & {\scriptsize{}0.13} & {\scriptsize{}600} & {\scriptsize{}Sato et al. \citeyearpar{Sato2014}}\tabularnewline
{\scriptsize{}Moon, Maria} & {\scriptsize{}-} & {\scriptsize{}0.22} & {\scriptsize{}-0.026} & {\scriptsize{}24} & {\scriptsize{}2.1} & {\scriptsize{}0.05} & {\scriptsize{}-} & {\scriptsize{}-} & {\scriptsize{}-} & {\scriptsize{}0.07} & {\scriptsize{}600} & {\scriptsize{}Sato et al. \citeyearpar{Sato2014}}\tabularnewline
{\scriptsize{}Deimos} & {\scriptsize{}C} & {\scriptsize{}0.079} & {\scriptsize{}-0.29} & {\scriptsize{}16} & {\scriptsize{}1.65} & {\scriptsize{}0.068} & {\scriptsize{}-} & {\scriptsize{}-} & {\scriptsize{}-} & {\scriptsize{}0.067} & {\scriptsize{}540} & {\scriptsize{}Thomas et al. \citeyearpar{1996Icar..123..536T}}\tabularnewline
{\scriptsize{}Phobos} & {\scriptsize{}C} & {\scriptsize{}0.07} & {\scriptsize{}-0.08} & {\scriptsize{}22} & {\scriptsize{}4.0} & {\scriptsize{}0.05} & {\scriptsize{}-} & {\scriptsize{}-} & {\scriptsize{}-} & {\scriptsize{}0.056} & {\scriptsize{}540} & {\scriptsize{}Simonelli et al. \citeyearpar{1998Icar..131...52S}}\tabularnewline
{\scriptsize{}9P/Tempel 1} & {\scriptsize{}JFC} & {\scriptsize{}0.039} & {\scriptsize{}-0.49} & {\scriptsize{}16} & {\scriptsize{}(1.0)} & {\scriptsize{}(0.01)} & {\scriptsize{}-} & {\scriptsize{}-} & {\scriptsize{}-} & {\scriptsize{}0.056} & {\scriptsize{}550} & {\scriptsize{}Li et al. \citeyearpar{2007Icar..187...41L}a}\tabularnewline
{\scriptsize{}19P/Borrelly} & {\scriptsize{}JFC} & {\scriptsize{}0.057} & {\scriptsize{}-0.43} & {\scriptsize{}22} & {\scriptsize{}(1.0)} & {\scriptsize{}(0.01)} & {\scriptsize{}-} & {\scriptsize{}-} & {\scriptsize{}-} & {\scriptsize{}0.072} & {\scriptsize{}660} & {\scriptsize{}Li et al. \citeyearpar{2007Icar..188..195L}b}\tabularnewline
{\scriptsize{}67P/C-G} & {\scriptsize{}JFC} & {\scriptsize{}0.032} & {\scriptsize{}-0.42} & {\scriptsize{}28} & {\scriptsize{}2.25} & {\scriptsize{}0.061} & {\scriptsize{}-} & {\scriptsize{}-} & {\scriptsize{}87} & {\scriptsize{}0.067} & {\scriptsize{}649} & {\scriptsize{}Fornasier et al. \citeyearpar{2015arXiv150506888F}}\tabularnewline
{\scriptsize{}81P/Wild 2} & {\scriptsize{}JFC} & {\scriptsize{}0.038} & {\scriptsize{}-0.52} & {\scriptsize{}27} & {\scriptsize{}(1.0)} & {\scriptsize{}(0.01)} & {\scriptsize{}-} & {\scriptsize{}-} & {\scriptsize{}-} & {\scriptsize{}0.063} & {\scriptsize{}647} & {\scriptsize{}Li et al. \citeyearpar{2009Icar..204..209L}}\tabularnewline
{\scriptsize{}103P/Hartley 2} & {\scriptsize{}JFC} & {\scriptsize{}0.036} & {\scriptsize{}-0.46} & {\scriptsize{}15} & {\scriptsize{}(1.0)} & {\scriptsize{}(0.01)} & {\scriptsize{}-} & {\scriptsize{}-} & {\scriptsize{}-} & {\scriptsize{}0.045} & {\scriptsize{}625} & {\scriptsize{}Li et al. \citeyearpar{2013Icar..222..559L}}\tabularnewline
 &  &  &  &  &  &  &  &  &  &  &  & \tabularnewline
\hline 
\end{tabular}
\par\end{centering}

\end{sidewaystable}

\par\end{center}

One of few bodies that allows further comparison to our update Hapke
parameters is (2867) Steins. (2867) Steins is an E-type, a type of
asteroid that generally presents a strong opposition effect \citep{2003Icar..166..276B}.
This body was one of the Rosetta asteroid targets and was observed
with the OSIRIS camera in similar conditions to (21) Lutetia, having
a phase angle that ranged from 0.36 to 141 degrees and varying pixel
resolution.\citet{2012Icar..221.1101S} used the latest Hapke model
to obtain global solution for all shape model facets that were corregistered
to rendered images. Then, they fixed all parameters, leaving just
one for variation. Hence they obtained residual maps of Hapke parameters
for all asteroid observed surface. The CBOE is quite small but it
improves the modeling at the opposition regime. Aside the surface
porosity value, we find little resemblance of the parameters between
both bodies.

\section{Summary and Conclusion}

We examined images taken in 18 filters by the NAC and WAC cameras
of the imaging system OSIRIS on-board Rosetta of the Baetica region
and surroundings. The region contains a cluster of superimposed craters
called NPCC, where the larger crater walls, called Gallicum Labes,
Danuvius Labes, Sarnus Labes, Low Corduba and High Corduba present
differences in morphology, texture and boulder distribution \citep{2012P&SS...66...96T}.
Aiming to characterize any spectro-photometrical variation among the
walls and concerning to investigate the effect of any space weathering
process, we undertook a disk-resolved photometric analysis. We applied
the latest Hapke model and we were able to draw a spatial distribution
map of the parameters. The major results are summarized as:
\begin{enumerate}
\item We observe a spectral dichotomy between Gallicum Labes-Low Corduba
and Danuvius Labes-Sarnus Labes-High Corduba. The Gallicum Labes and
Low Corduba are around 40\% redder than average spectrum ($\bar{\gamma}(5^{\circ})=3.5\pm0.55\%\cdot\mu m^{-1}$,
$\bar{\gamma}(20^{\circ})=2.9\pm0.38\%\cdot\mu m^{-1}$), while Danuvius-Sarnus
Labes and High Corduba are 15-40\% bluer. The variegation, otherwise,
is no higher than 8\% for Gallicum-Sarnus Labes. Sarnus has a more
intriguing behavior, since it shares characteristics of Gallicum and
Danuvius Labes, it is spectrally closer to the latter, but it has
similar normal albedo to Gallicum Labes. If compared to Psyche Crater
in (433) Eros \citep{2001M&PS...36.1617C}, the correlation is opposite,
the variegation is larger (32-40\%), but spectrally, the dark material
at the bottom of the crater and the bright material on the walls,
show no more than 4-8\% of difference. Thus, we have confirmed that
the variegation and spectral differences between crater walls are
significant, as hinted by the small phase angle images of (21) Lutetia.
We emphasize that the compositional nature of (21) Lutetia is not
accessed by our spectral range or photometric analysis. Lutetia shows
a featureless spectra with varying spectral slope. \citet{2012P&SS...66...23B}
have previously concluded that the surface composition is ambiguous,
possibly a chondritic mixture of carbonaceous and enstatite. 
\item  Gallicum Labes and Low Corduba have the brightest wall ($\bar{A}_{0}=0.214\pm0.005$)
with accentuated asymmetric factor ($\bar{g}_{sca}=-0.35\pm0.5$)
and with sharper SHOE ($\bar{B}s_{0}=0.81\pm0.03$, $h_{s}=0.029\pm0.002$)
and broader CBOE ($\bar{B}c_{0}=0.10\pm0.004$, $h_{c}=0.07\pm0.002$).
Danuvius Labes and High Corduba, on the other hand, are darker ($\bar{A}_{0}=0.199\pm0.005$),
slightly less back-scatter ($\bar{g}_{sca}=-0.30\pm0.1$) with narrower
CBOE ($\bar{B}c_{0}=0.065\pm0.002$, $h_{c}=0.056\pm0.003$) with
average SHOE ($\bar{B}s_{0}=0.83\pm0.02$, $h_{s}=0.043\pm0.001$).
The average roughness slope ($\bar{\theta}=11-12^{\circ}$) for all
the computed areas indicates a smooth terrain.
\item We have not detected any expressive global phase reddening or bluing
on Baetica ($-0.04\pm0.045\%\cdot\mu m^{-1}deg^{-1}$) between phase
angles of $5^{\circ}$ and $20^{\circ}$. The phase angle range between
the two sets of images is small to uncover the effect . Locally, we
find hint of a slight phase bluing on High Corduba ($-0.053/-0.066\pm0.045\%\cdot\mu m^{-1}deg^{-1}$).
The Highland is connected to a smooth and and darker terrain with
different grain distribution than Gallicum Labes.
\item Measurements with VIRTIS \citep{2011Sci...334..492C} and MIRO \citep{2012P&SS...66...31G}
instruments reported very low thermal inertia of 20-30 $J\cdot m^{-2}\cdot K^{-1}\cdot s^{-\frac{1}{2}}$
or even less. The range is consistent with highly porous upper regolith
layer of 1-3 cm thick. A roughness modeling by Keihm et al. (2012)
affirms that a surface composed of 50\% of unresolved craters of 1
cm or larger are necessary to correct the MIRO-VIRTIS measurement
offset. Therefore, the thermal data points to a porous regolith layer
with an amount of sub-pixel roughness. Our Hapke modeling of 85\%
porosity and $11.5^{\circ}$ of average roughness slope, however,
is partially consistent with this interpretation. Our $\bar{\theta}$
indicates a smooth surface, but it is not representative of the whole
Lutetia surface, since we model only Baetica and part of the Etruria
region. However, our global Hapke (1993) modeling returns $\bar{\theta}=29^{\circ}$
, a much rougher and compatible surface. This parameter is influenced
by the inclusion of the rougher Narbonensis region, which a preliminary
modeling indicated $\bar{\theta}$ranging from $20^{\circ}$ to $30^{\circ}$.
\item We compared our Hapke parameters with laboratory samples and other
disk-resolved small bodies of the Solar System. None of the bodies
completely match Baetica, although the global Hapke (1993) parameters
of Lutetia match some S-type asteroids. However, as Lutetia spectrum
diverges from this taxonomic type, the similarities may only be related
to the state of the regolith. Comparing with laboratory, we conclude
that Cobalt Oxide (700 nm) and Red Iron Oxide (550 nm) are the best
analogs for the morphological characteristics of the regolith in Baetica.
\item The presence of the CBOE is ambiguous. The addition of the mechanism
does not improve the fitting. We find that the SHOE parameters are
robust for the tests conducted with and without CBOE. We thus conclude
that shadow-hiding must be the dominant opposition effect mechanism,
with an average HWHM of $4.6^{\circ}$.
\item Even after photometric correction with two different approaches, Gades
and Corduba remains collectively different from the surroundings on
different ways. Gallicum Labes and Low Corduba are redder and composed
of bright material exposed by nearby perturbation or impact. The Hapke
parameters point to a porous regolith distribution of fine grains.
The morphology in the region points to landslides with no further
signs of craterization over it. To reinforce this assumption, we observe
the bright structure J in Etruria. This structure is few pixels wide
($\sim4$ km) and starts from a small crater and comes down to a deeper
depression, as the filaments also found in Gallicum. However, we cannot
certainly state that Gallicum is composed of a fresher material that
has not suffered further space weathering or just a different composition.
\end{enumerate}

\subsubsection*{Acknowledgements}

\emph{\footnotesize{}The authors thanks CNPq, process no.402085/2012-4,
for the support. FAPERJ and CAPES are also aknowledged for diverse
grants and fellowships to D.L. and P.H.H.}{\footnotesize \par}

\emph{\footnotesize{}OSIRIS was built by a consortium of the Max-Planck-
Institut f�r Sonnensystemforschung, G�ttingen, Germany, CISAS\textendash University
of Padova, Italy, the Laboratoire d\textquoteright Astrophysique de
Marseille, France, the Instituto de Astrof�sica de Andalucia, CSIC,
Granada, Spain, the Research and Scientific Support Department of
the European Space Agency, Noordwijk, The Netherlands, the Instituto
Nacional de T�cnica Aeroespacial, Madrid, Spain, the Universidad Polit�chnica
de Madrid, Spain, the Department of Physics and Astronomy of Uppsala
University, Sweden, and the Institut f�r Datentechnik und Kommunikationsnetze
der Technischen Universit�t Braunschweig, Germany. The support of
the national funding agencies of Germany (DLR), France (CNES), Italy
(ASI), Spain (MEC), Sweden (SNSB), and the ESA Technical Directorate
is gratefully acknowledged.}{\footnotesize \par}

\bibliographystyle{agu04}
\bibliography{lutetia}

\end{document}